\documentclass[reprint,amsmath,twocolumn,amssymb,prl,superscriptaddress]{revtex4}
\usepackage{graphicx}
\usepackage{dcolumn}
\usepackage{bm}
\usepackage[utf8]{inputenc}
\usepackage{color}
\usepackage{hyperref}
\usepackage{siunitx}
\usepackage{amsmath}
\usepackage{ulem}
\usepackage{float}
\definecolor{purple}{RGB}{128,0,128}

\newcommand{\ind}[2]{\raisebox{#1}{#2}}

\let\vec\mathbf

\begin{document}

\title{Electronic spectrum of twisted graphene layers under heterostrain}

\author{Loïc Huder}
\affiliation{Univ. Grenoble Alpes, CEA, INAC, PHELIQS, F-38000 Grenoble, France}
\author{Alexandre Artaud}
\affiliation{Univ. Grenoble Alpes, CEA, INAC, PHELIQS, F-38000 Grenoble, France}
\affiliation{Univ. Grenoble Alpes, CNRS, Inst. NEEL, F-38000 Grenoble, France}
\author{Toai Le Quang}
\affiliation{Univ. Grenoble Alpes, CEA, INAC, PHELIQS, F-38000 Grenoble, France}
\author{Guy Trambly de Laissardière}
\affiliation{Laboratoire de Physique Théorique et Modélisation, Université de Cergy-Pontoise-CNRS, F-95302 Cergy-Pontoise, France}
\author{Aloysius G. M. Jansen}
\affiliation{Univ. Grenoble Alpes, CEA, INAC, PHELIQS, F-38000 Grenoble, France}
\author{Gérard Lapertot}
\affiliation{Univ. Grenoble Alpes, CEA, INAC, PHELIQS, F-38000 Grenoble, France}
\author{Claude Chapelier}
\affiliation{Univ. Grenoble Alpes, CEA, INAC, PHELIQS, F-38000 Grenoble, France}
\author{Vincent T. Renard}
\affiliation{Univ. Grenoble Alpes, CEA, INAC, PHELIQS, F-38000 Grenoble, France}


\date{\today}



\begin{abstract}
We demonstrate that stacking layered materials allows a novel type of strain engineering where each layer is strained independently, which we call heterostrain. We combine detailed structural and spectroscopic measurements with tight-binding calculations to show that small uniaxial heterostrain suppresses Dirac cones and leads to the emergence of flat bands in twisted graphene layers (TGLs). Moreover, we demonstrate that heterostrain reconstructs much more severely the energy spectrum of TGLs than homostrain for which both layers are strained identically ; a result which should apply to virtually all van der Waals structure opening exciting possibilities for straintronics with 2D materials.
\end{abstract}

\maketitle

\begin{figure*}
\ind{3cm}{a}~\raisebox{-0.07cm}{\includegraphics[width=0.29\textwidth]{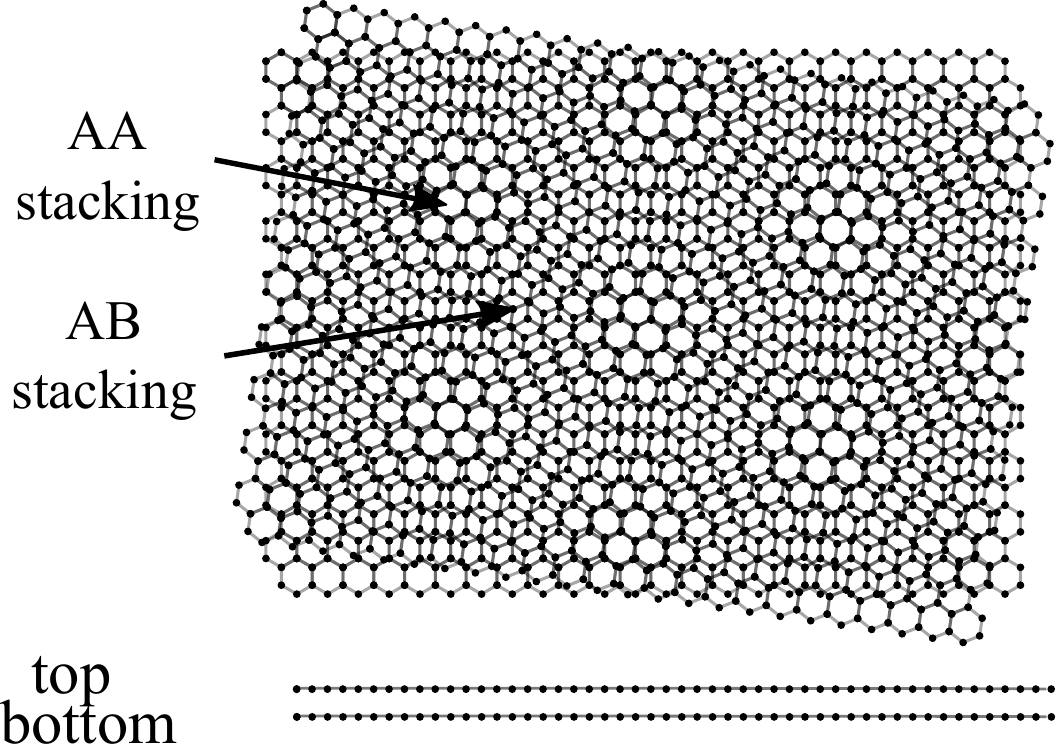}}
\ind{3cm}{b}~\includegraphics[width=0.32\textwidth]{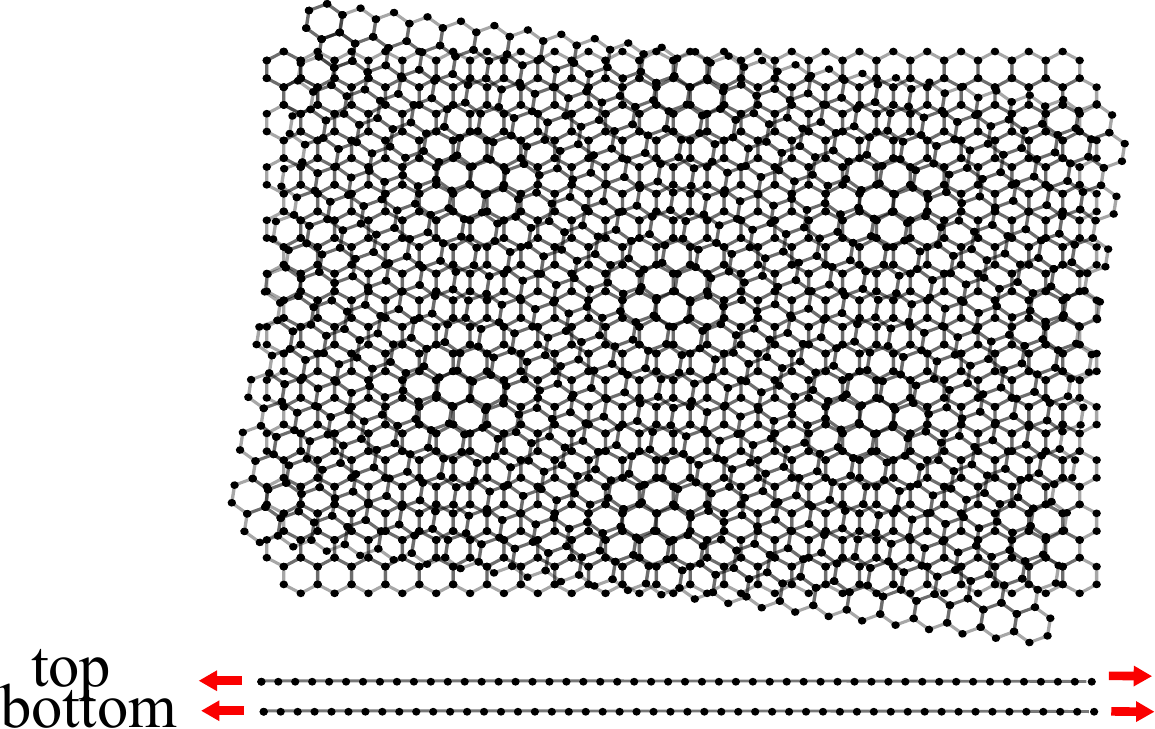}
\ind{3cm}{c}~\includegraphics[width=0.32\textwidth]{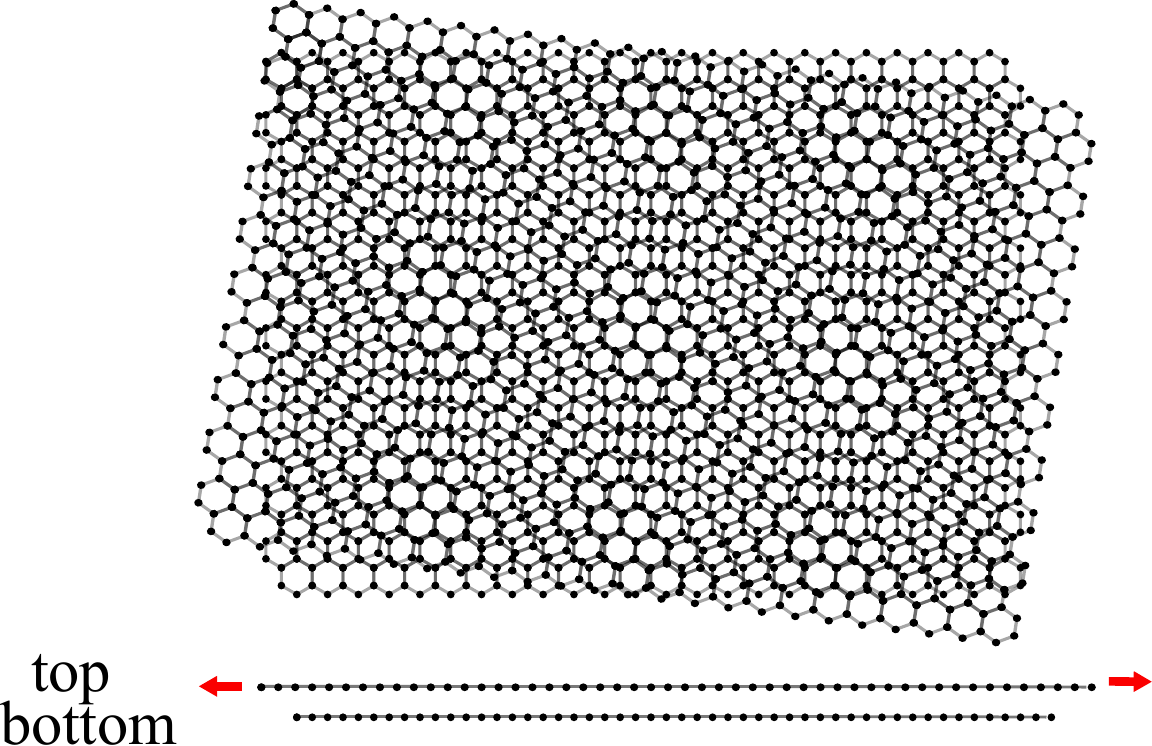}
\ind{5.5cm}{d}~\includegraphics[width=0.425\textwidth]{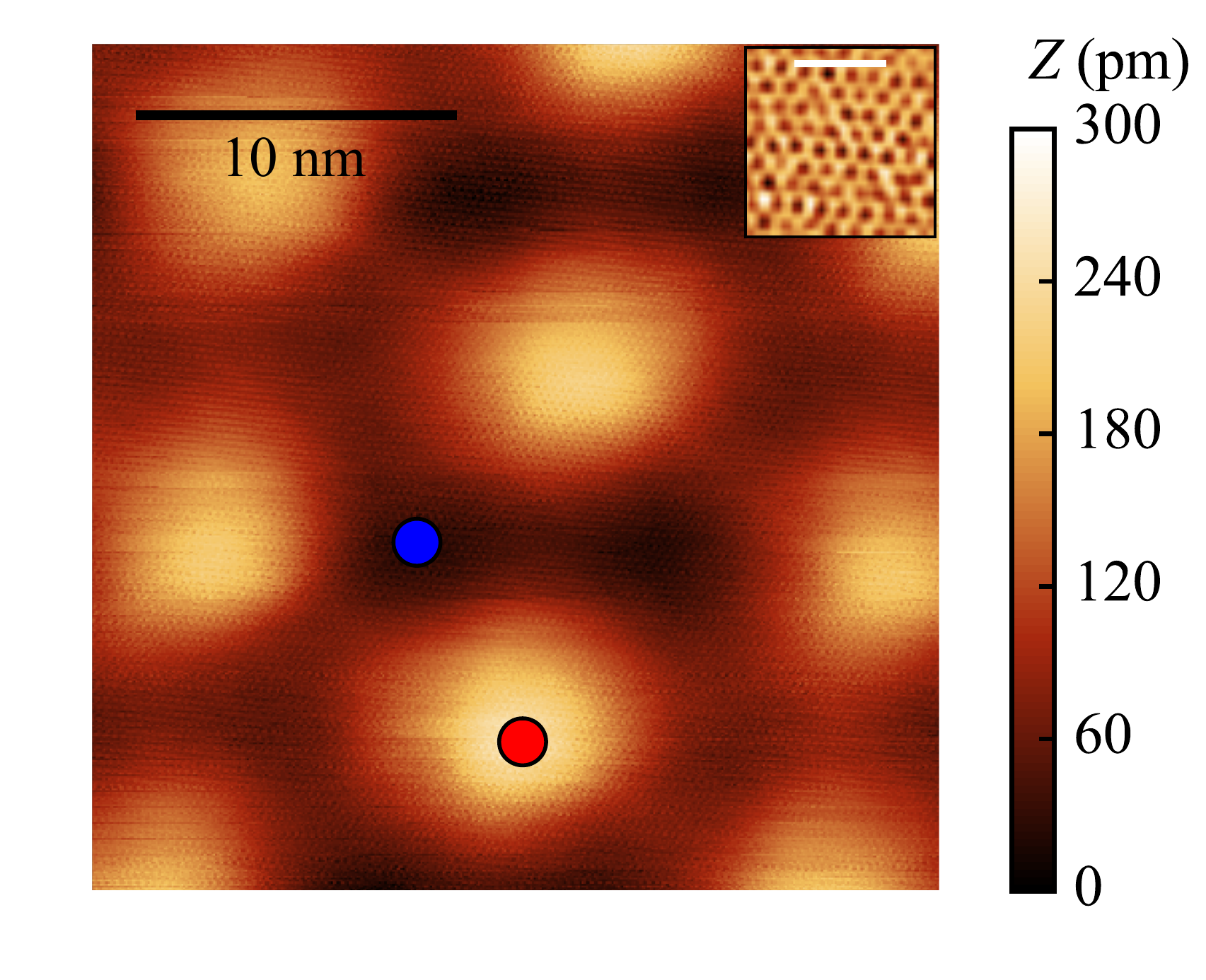}
\ind{5.5cm}{e}~\includegraphics[width=0.425\textwidth]{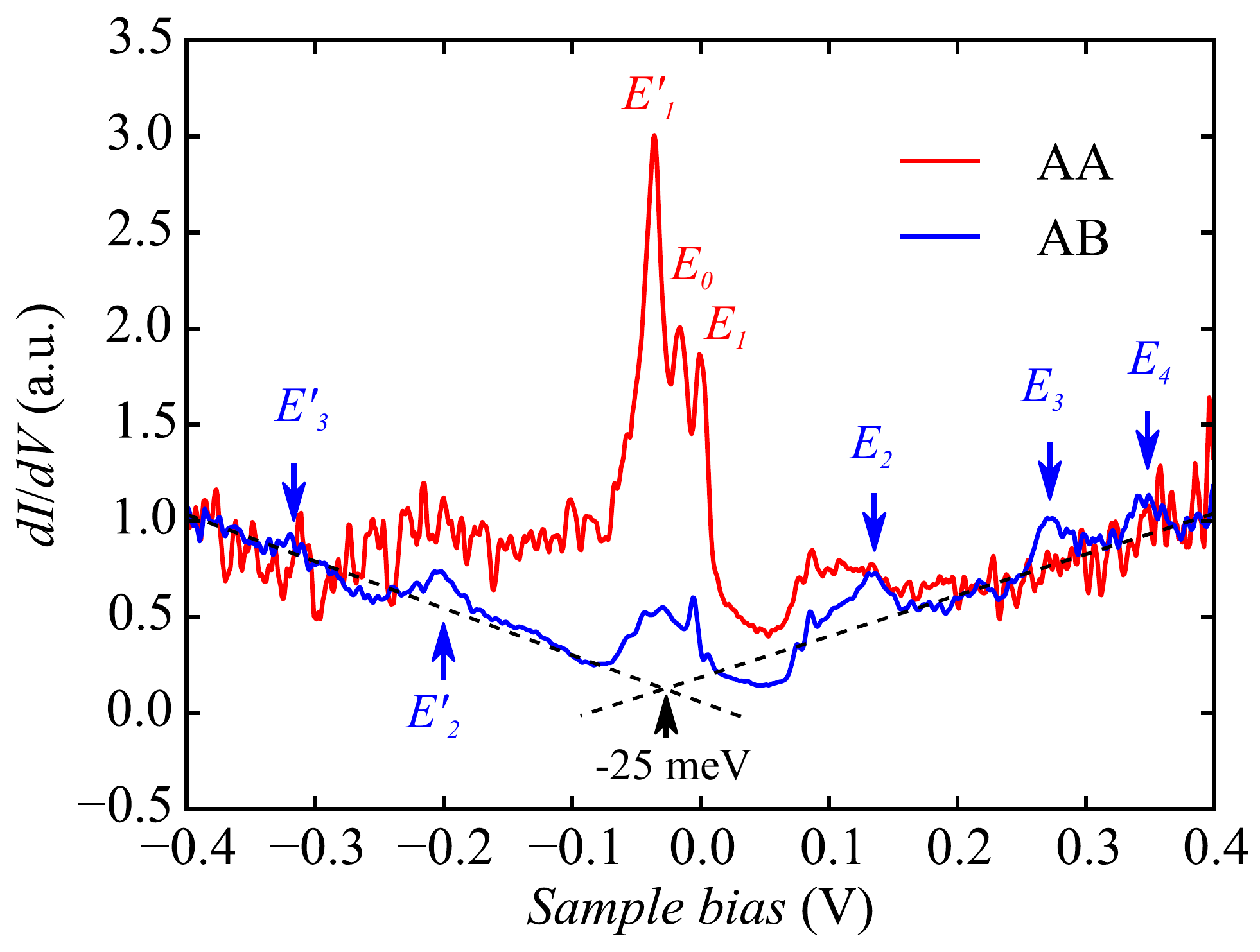}
\caption{\textbf{Strained twisted graphene layers.} a), b) and c) Sketch of TGLs without or with strain; top and side views. Red arrows in the side views indicate the presence of strain in each layer. a) TGLs without strain. b) TGLs with 10\% of uniaxial strain applied to both layers. c) TGLs with 10 \% of uniaxial strain applied to the top layer only. d) 26.4$\times$26.4~nm$^2$ topograph of twisted graphene layers showing the moiré pattern. The twist angle between the graphene layers was estimated to be $\theta$=\ang{1.25}. Bias voltage $V=-400$~mV and current set-point $I=50$~pA. Inset: Zoom of the image showing the honeycomb lattice of the carbon atoms in the top layer. The scale bar is 1~nm e) Differential conductance (dI/dV) recorded at the spots marked with colored dots in Fig.~\ref{f:stm_sts}a (red and blue dots for AA and AB regions respectively)  showing multiple resonances in AA regions (red arrows) and well resolved peaks at higher energy in AB regions (blue arrows). Bias voltage $V=-400$~mV and current set-point $I=300$~pA. The doping is estimated from the linear extrapolation of the high-energy density of states (dashed lines).
}
\label{f:stm_sts}
\end{figure*}

A variety of new electronic, optoelectronic and photovoltaic devices are produced by stacking different two-dimensional materials into van der Waals heterostructures.\cite{Geim2013,Novoselov2016a} Radical effects on the electronic properties can also be achieved by stacking identical materials into van der Waals homostructures. This is well illustrated by graphene bilayers\cite{Ohta2006} and transition metal dichalcogenide bilayers\cite{Mak2010} which are drastically different from their monolayer counterpart. The properties of van der Waals structures and device not only depend on the choice of materials but also on the details of the stacking which offers additional degrees of freedom.\cite{Novoselov2016a} The rotation angle between the layers has been exploited to tune the van Hove singularities in twisted graphene layers (TGLs),\cite{VanHove1953,LopesdosSantos2007,Li2009a,TramblydeLaissardiere2010,Bistritzer2011,LopesdosSantos2012,TramblydeLaissardiere2012,Brihuega2012,Wong2015} to achieve the first experimental observation of the Hofstadter butterfly in graphene on h-BN\cite{Dean2013,Ponomarenko2013} and to uncover correlated insulator and superconducting states in magic angle twisted graphene layers.\cite{Hererro2018a,Hererro2018b} It has also allowed to monitor the optical responses of transition metal dichalcogenides homo-\cite{liu2014} and heterostructures.\cite{Nayak2017} Strain, originating from external stress\cite{Yan2013,Nguyen2015,He2016,Huang2016} or from interlayer interactions,\cite{Woods2014,jung2015,Kumar2015} is another degree of freedom.

In the present work we have studied the effect of homogeneous heterostrain on electronic properties of TGLs. Here, the layers experience different and independent homogeneous in-plane strain, a situation which has no equivalent in bulk materials where covalent bonds link atoms on both sides of the junction. Heterostrain is possible in 2D stacks because independent deformations on each side of the interface are allowed by the weak interlayer van der Waals bonding. The effect of strain on the structure of TGLs is sketched in Fig.~\ref{f:stm_sts}. In absence of strain (Fig.~\ref{f:stm_sts}a), the TGLs shows a typical moiré pattern resulting from the superposition of the two atomic lattices. With uniaxial strain applied identically to both layers (homostrain), the moiré is slightly deformed (Fig.~\ref{f:stm_sts}b). The moiré appears to be much more affected by a similar deformation of the top layer only (heterostrain is illustrated in Fig.~\ref{f:stm_sts}c). This magnifying effect of the moiré has been shown recently to allow for the determination of heterostrain\cite{Falko2014, Artaud2016, Summerfield2016, Jiang2017} and a strong influence on the moiré physics was predicted.\cite{Falko2014} Since TGLs, like other van der Waals structures, inherit their band structure from the moiré potential one may wonder whether this apparently large influence of heterostrain on the moiré has also strong implications for their electronic properties.

Figures~\ref{f:stm_sts}d and e demonstrate that this is indeed the case. Figure~\ref{f:stm_sts}d shows a constant-current Scanning Tunnelling Microscope (STM) image of a TGLs structure on SiC (000$\overline{1}$) obtained using the growth recipe described in Ref.~\onlinecite{Kumar2016}. The image shows a moiré pattern similar to those depicted in Figs.~\ref{f:stm_sts}a, b and c. The contrast results from alternation of AA-stacked regions, where the layers are in perfect registry (bright regions in the STM image) and of AB-stacked regions where the layer stacking is similar to graphite (dark regions in the image). Figure~\ref{f:stm_sts}e shows the differential conductance dI/dV which is proportional to the local density of states LDOS. The differential conductance in AA and AB regions was recorded as a function of the applied bias voltage $V$ using phase sensitive detection (2~mV modulation of  the bias voltage at 263~Hz. See suplementary section A for more details about reproducibility and spatial dependence of the data). The measurements were performed at T = 50 mK. The most striking feature is the presence of a group of resonances ($E'_1$, $E_0$, $E_1$) located near zero energy which are much more intense in AA regions. The linear extrapolation of the LDOS at high energy (dashed line) allows to deduce that this group of states is centred around $E=-25$~meV which corresponds to the typical doping of graphene grown on the carbon-face of SiC\cite{Sun2010}. These three resonances are surprising since TGLs usually present only two resonances (van Hove singularities) flanking the Dirac point\cite{VanHove1953,LopesdosSantos2007,Li2009a,TramblydeLaissardiere2010,Bistritzer2011,LopesdosSantos2012,TramblydeLaissardiere2012,Brihuega2012,Wong2015} and we will show that this qualitative difference can be attributed to the modification of the band structure by a small heterostrain.  

\begin{figure*}
  \fcolorbox{white}{white}{
  \begin{minipage}{0.295\textwidth}
  \ind{3.5cm}{a}~\includegraphics[height=3.7cm]{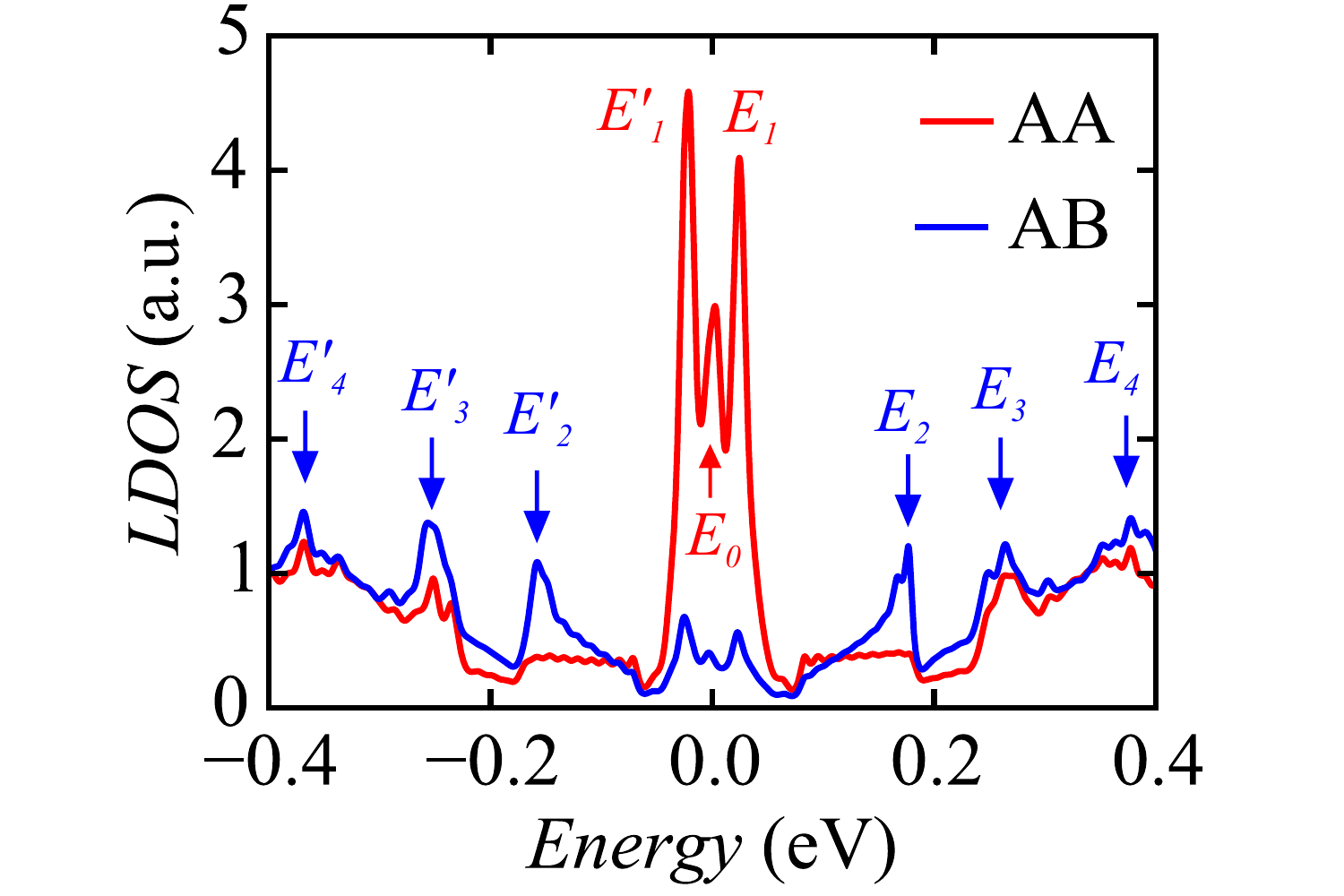}
   \includegraphics[height=6.5cm]{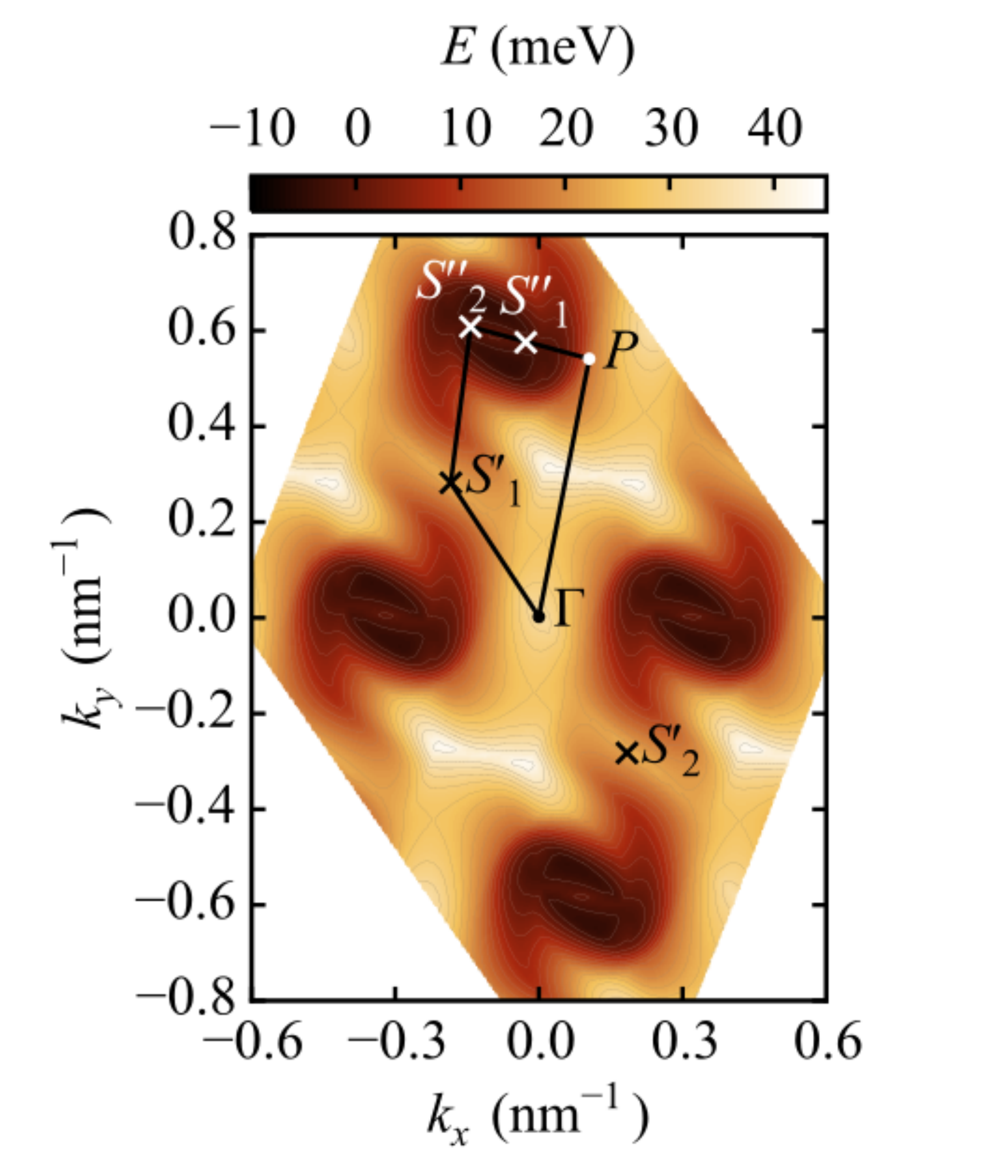}
  \includegraphics[height=3.7cm]{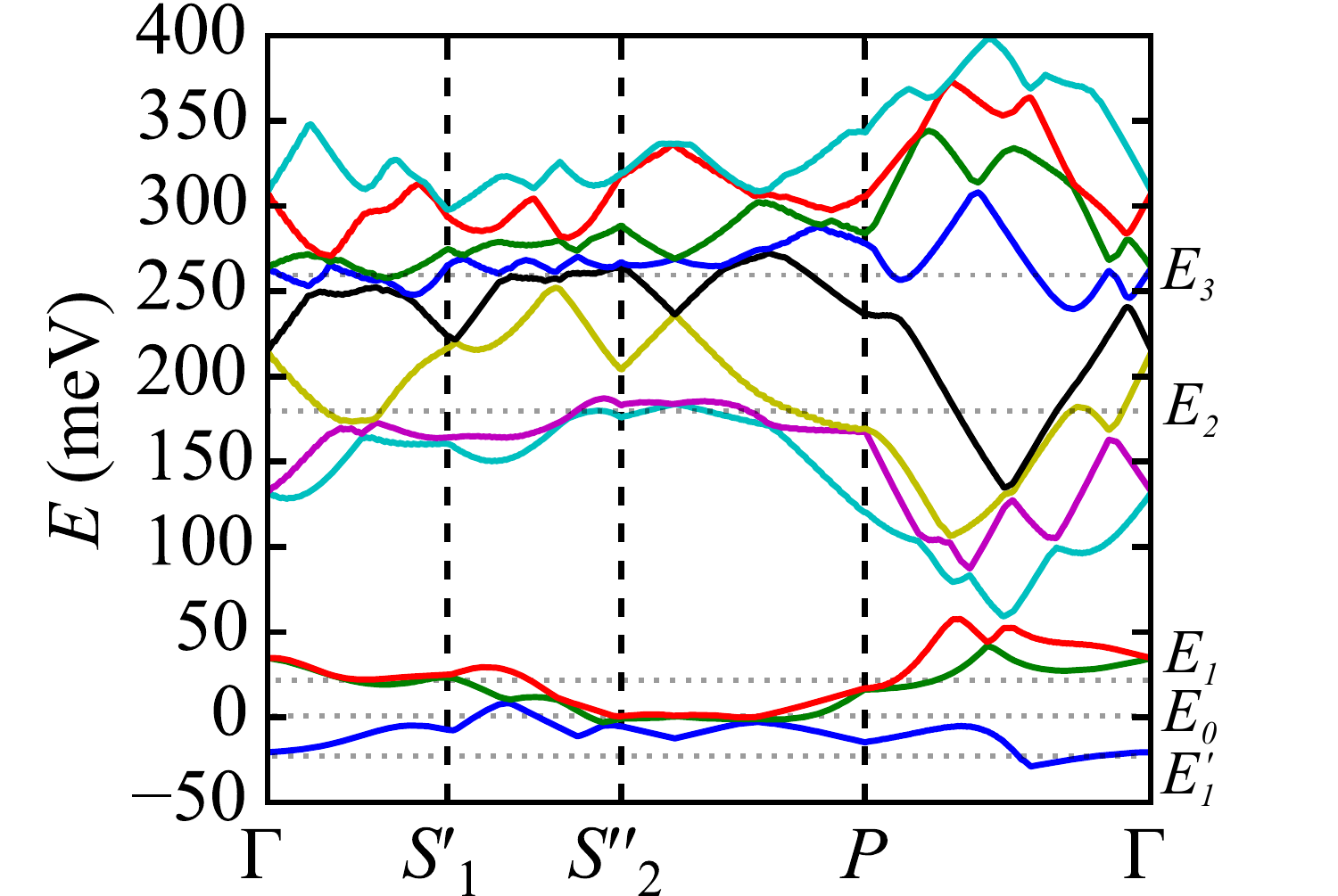}
  \end{minipage}
  }
  \fcolorbox{white}{white}{
  \begin{minipage}{0.295\textwidth}
  \ind{3.5cm}{b}~\includegraphics[height=3.7cm]{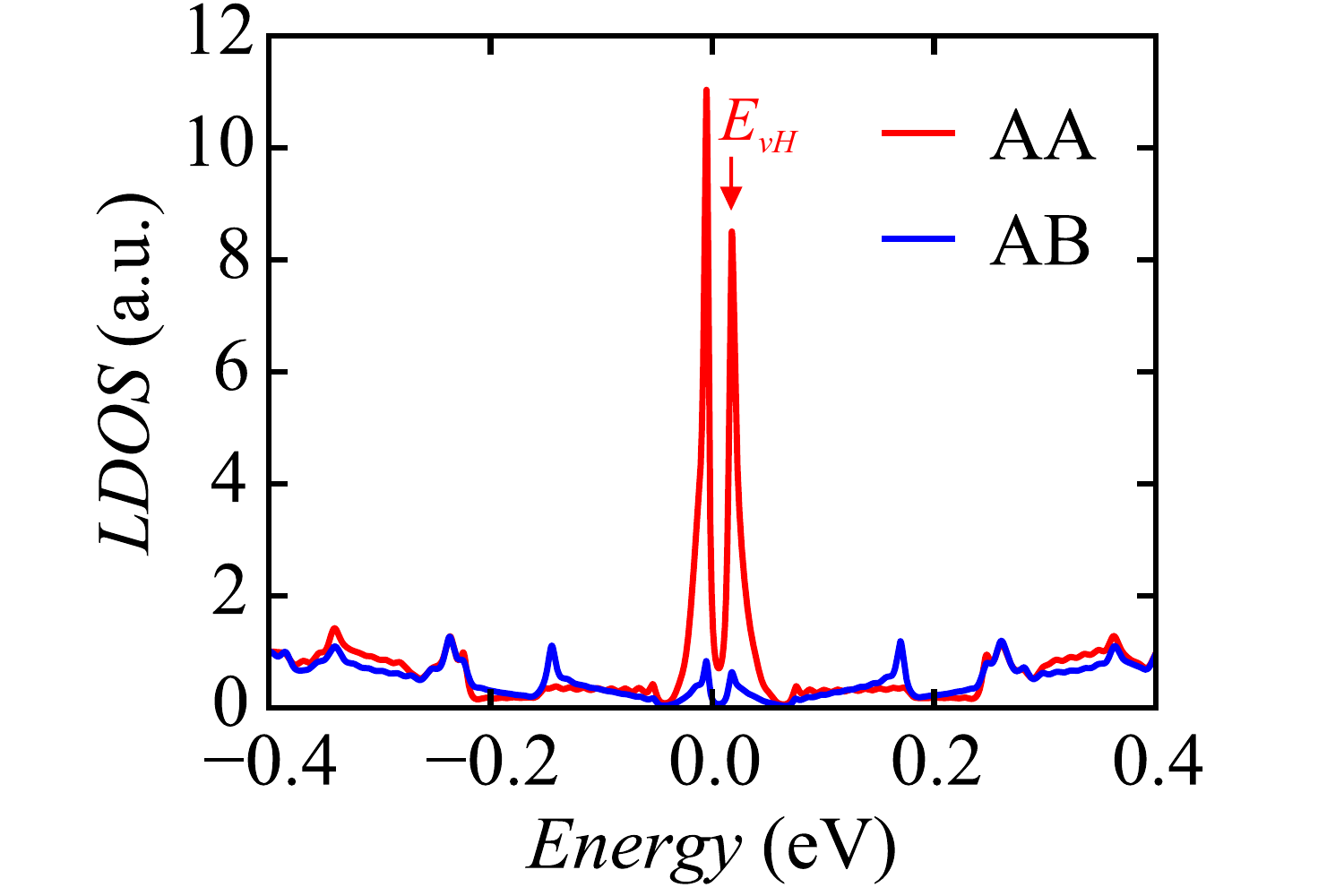}
   \includegraphics[height=6.5cm]{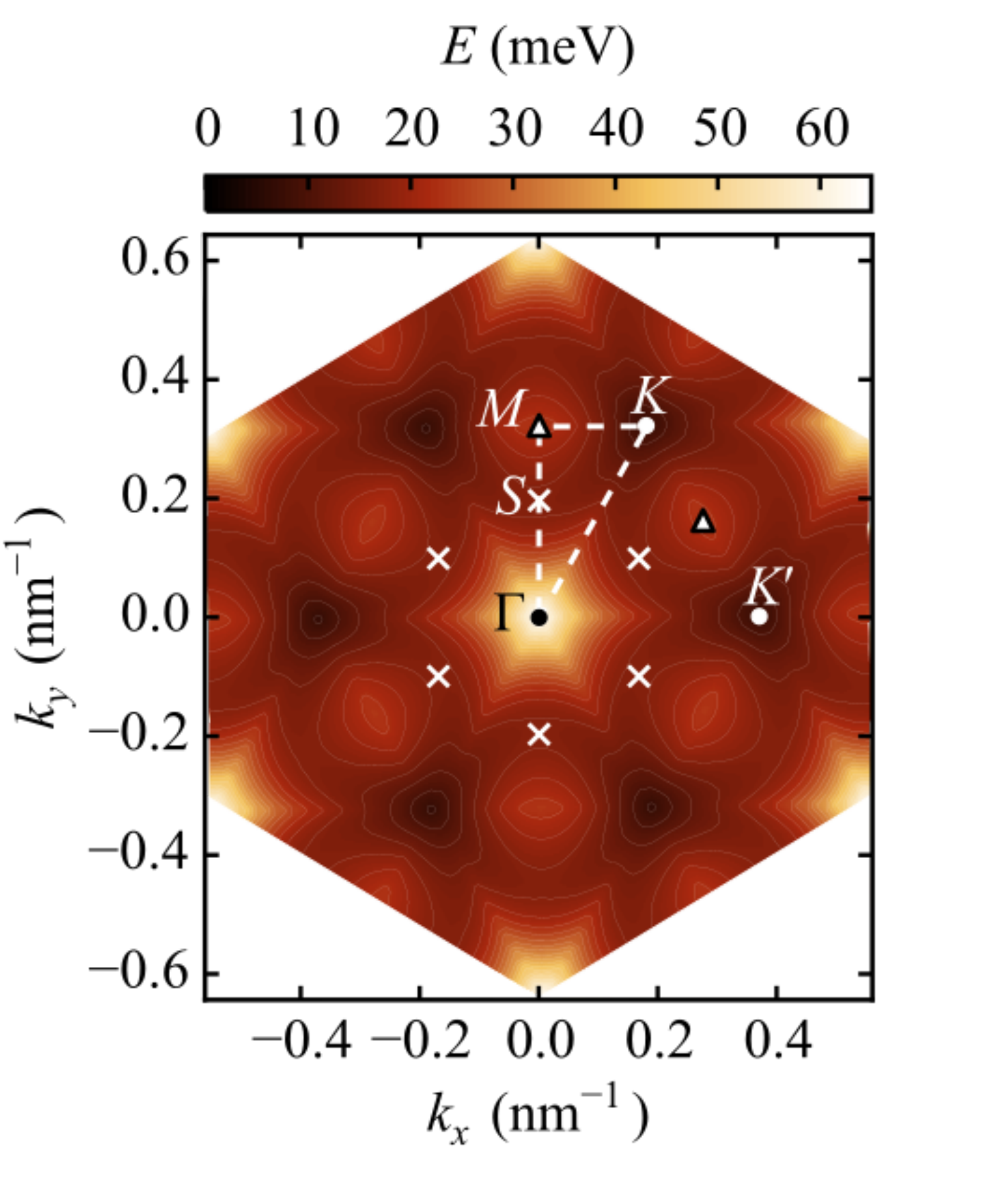}
   \includegraphics[height=3.7cm]{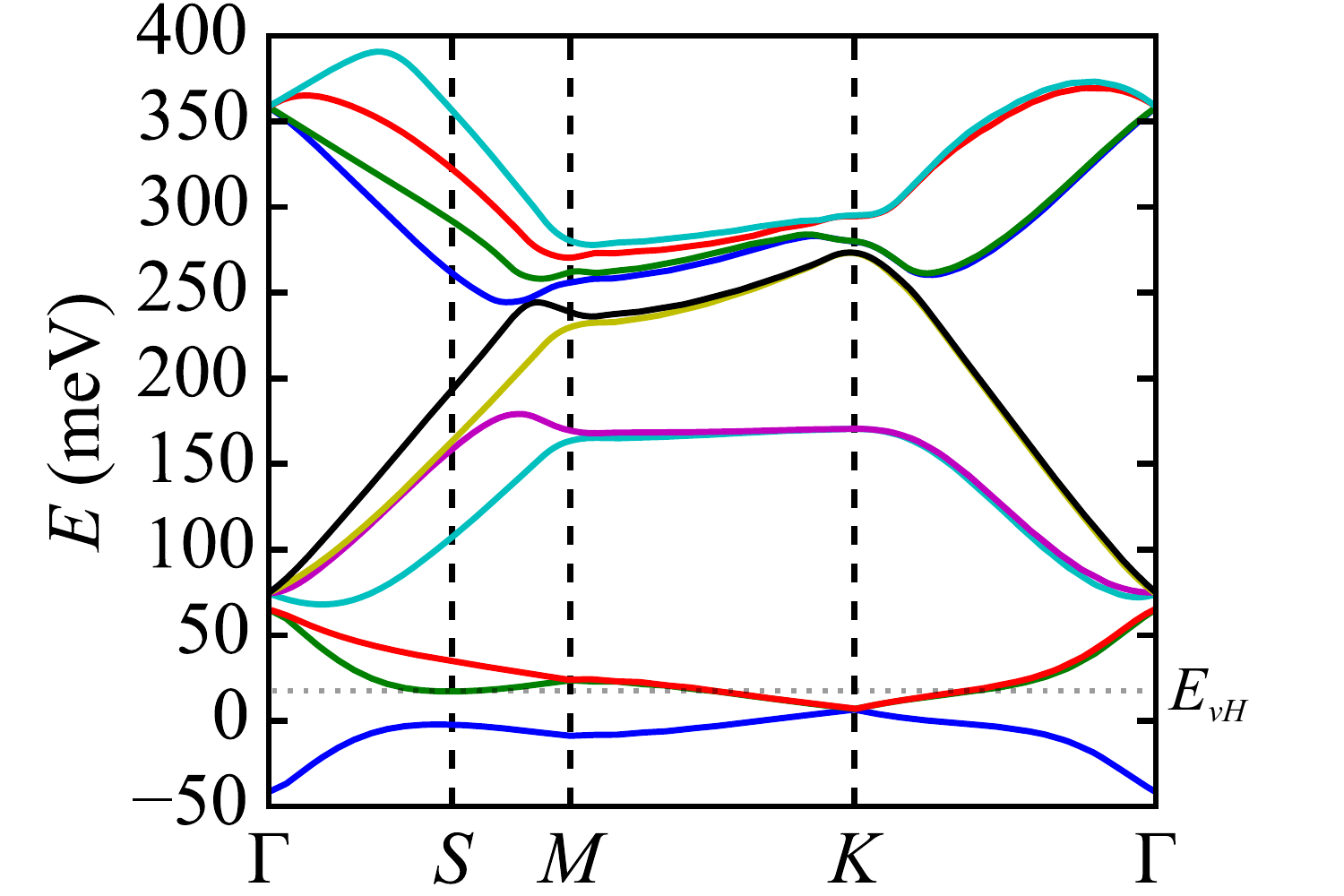}
   \end{minipage}
   }
   \fcolorbox{white}{white}{
  \begin{minipage}{0.295\textwidth}
  \ind{3.5cm}{c}~\includegraphics[height=3.7cm]{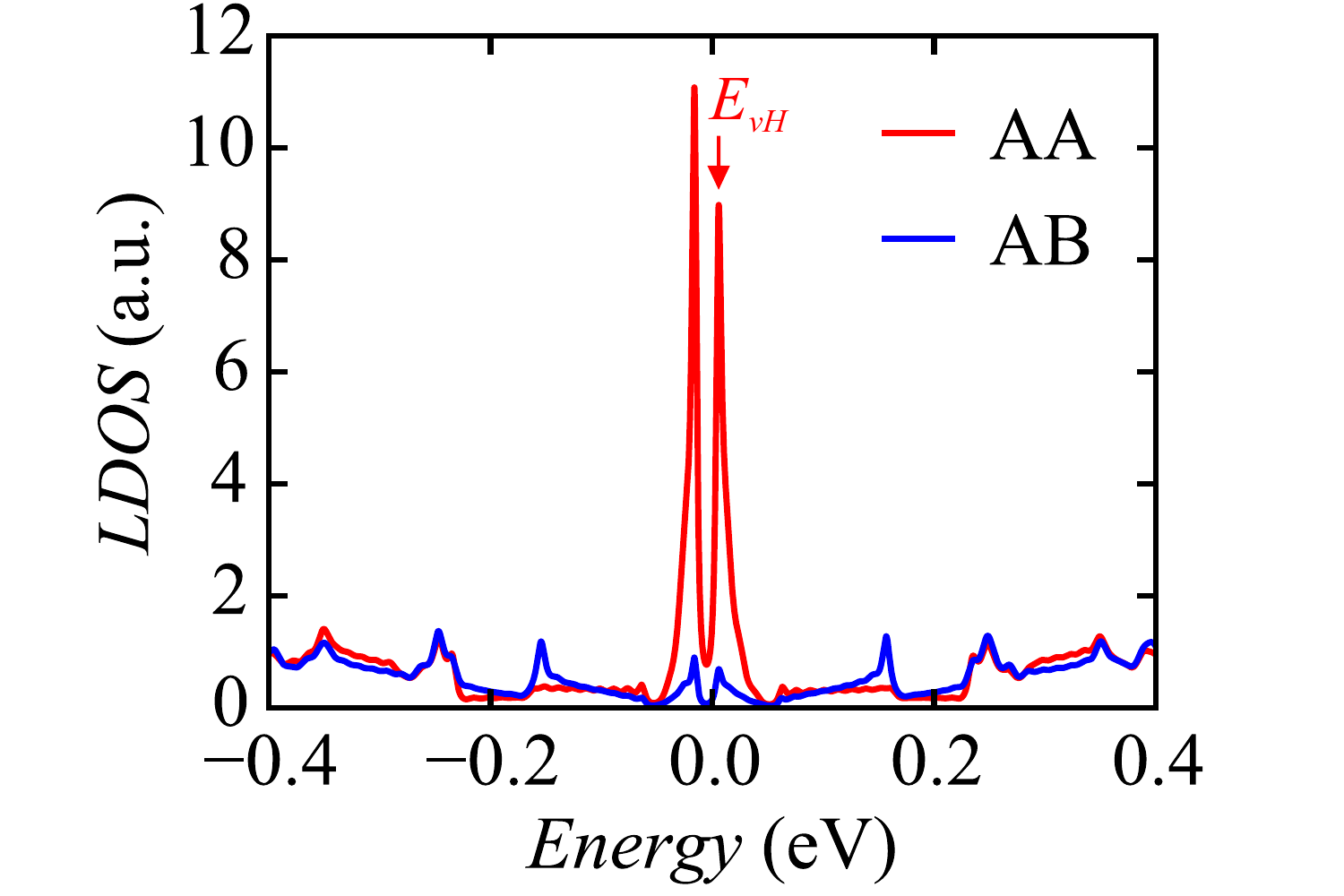}
   \includegraphics[height=6.5cm]{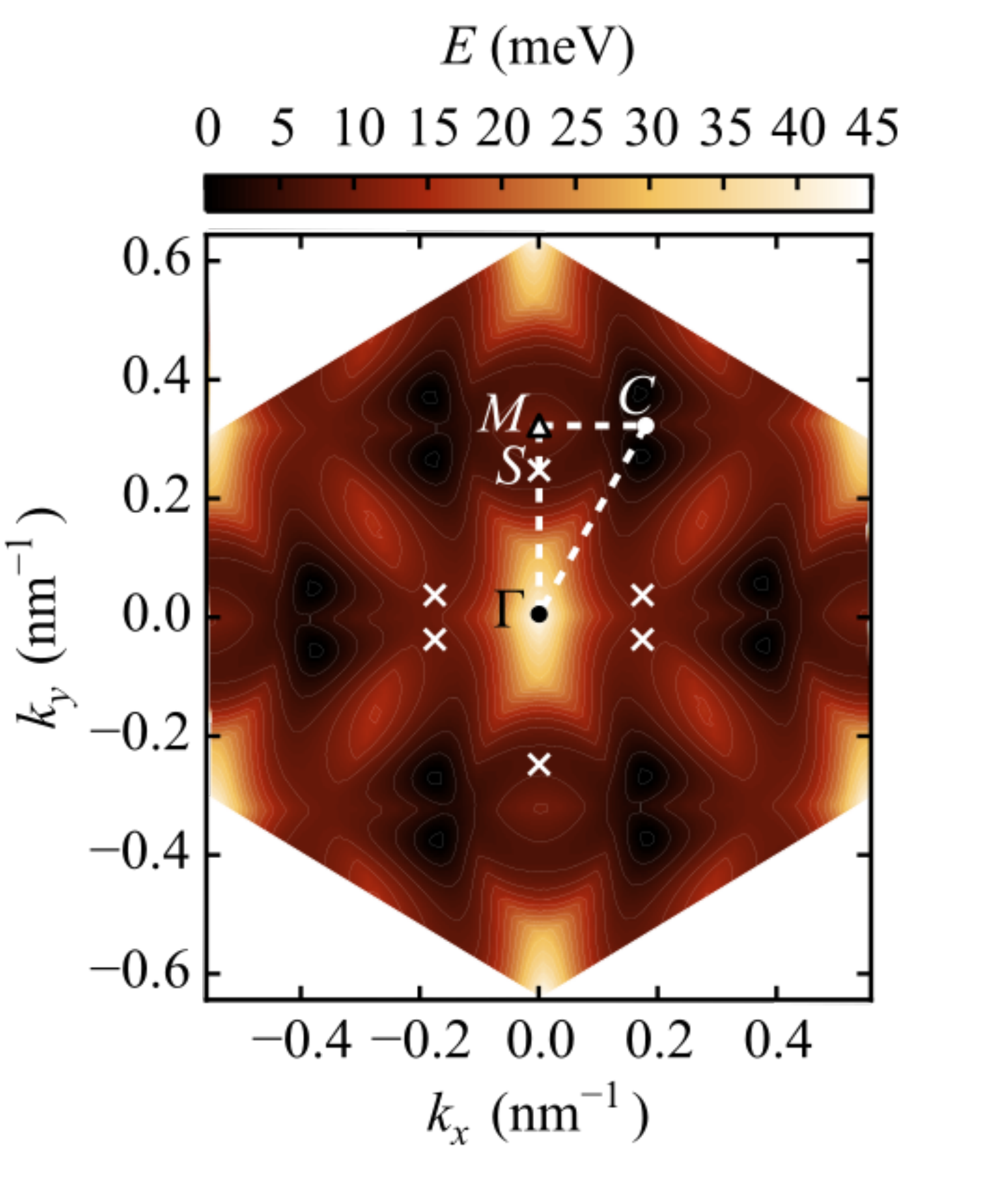}
   \includegraphics[height=3.7cm]{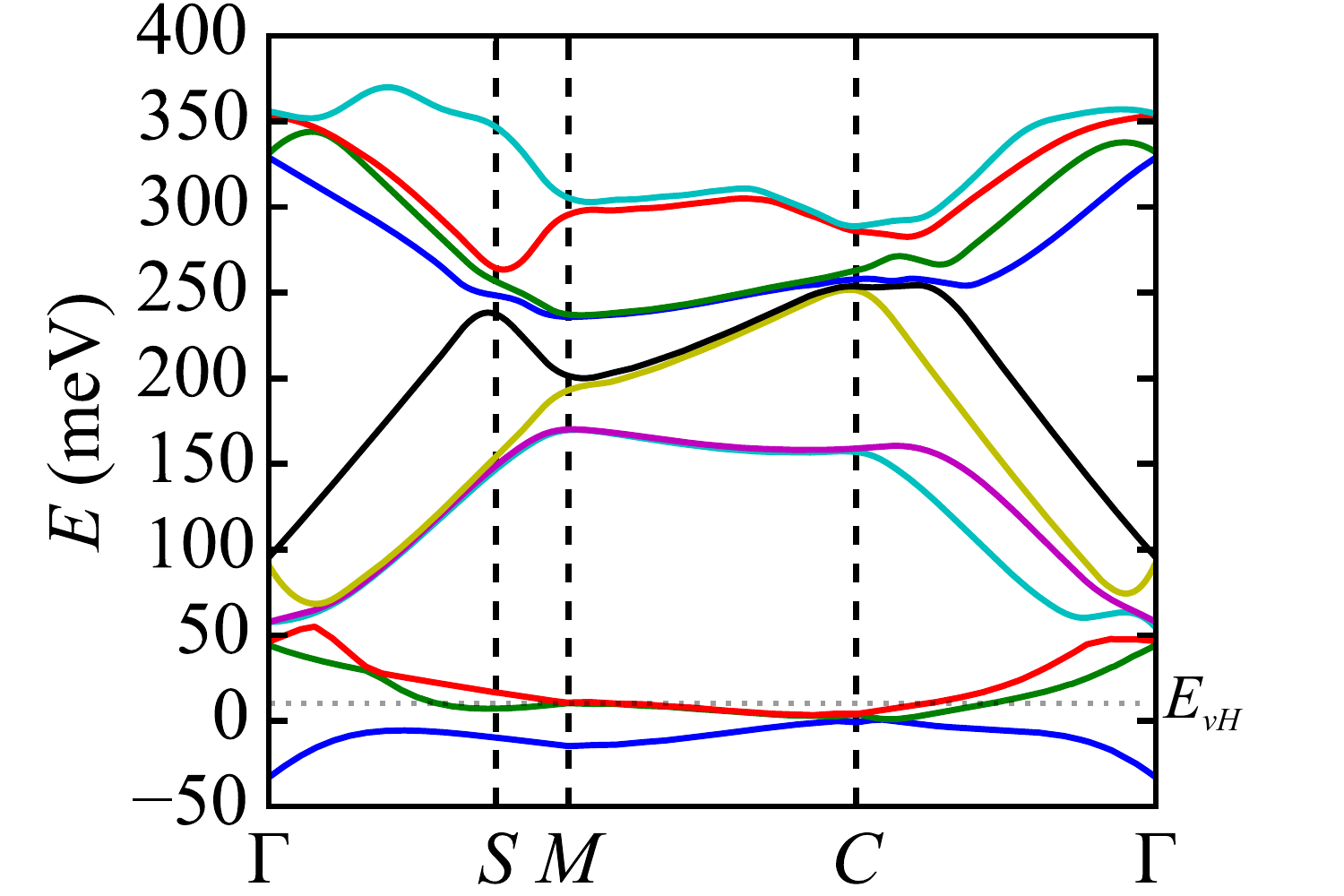}
   \end{minipage}
   }
\caption{\textbf{Tight binding calculations with and without strain.} The figure is organized in columns. In each column the top panel shows the LDOS in AA and AB regions. The middle panel shows the calculated energy map of the first band above $E_F$ (Crosses indicate saddle points). Lines correspond to the trajectory of the cuts in the band structure presented in the lower panel which shows the dispersion of the valence band and the first ten bands above $E_F$. a) Result for the structure with heterostrain of 0.35 \% corresponding to the experimental situation. b) Result for the unstrained structure with a twist angle close to the experimental situation. c) Result for the same structure as panel b) with an homostrain of $0.35~\%$. In the lower part of panel c), for simplicity, we have plotted cuts in $\Gamma C$ and $MC$ since the degeneracy at $K$ and $K'$ points is lifted by homostrain as pointed in Ref.~\onlinecite{Nguyen2015}. }
\label{f:tb}
\end{figure*}

A detailed Fourier analysis\cite{Artaud2016} of the image in Fig.~\ref{f:stm_sts}d  reveals that the layers are rotated by 1.25$^\circ$ and, more importantly, that one layer is slightly strained with respect to the other layer. This heterostrain consists in a combination of a small uniaxial heterostrain $\varepsilon_{uni}^{het}=0.35\% (\pm 0.03\%)$ applied along the horizontal direction in Fig.~\ref{f:stm_sts}d and an even smaller biaxial heterostrain $\varepsilon_{bi}^{het}=-0.06\% (\pm 0.005\%)$ (see supplemental material B for the detailed analysis). We note that possible local deformations are ignored by the Fourier analysis since it is performed on the entire image of Fig.~\ref{f:stm_sts}d. The above values correspond to a spatially averaged heterostrain. This small heterostrain originates from the pinning of the top layer at its boundaries during the growth. It is substantiated by lattice deformations at the boundary of the grain studied (see supplemental material C for more details).

We have investigated theoretically the influence of this small heterostrain on the electronic properties. Local densities of states are computed by recursion method in real space from a tight-binding Hamiltonian with Slater-Koster parameters for $p_z$ orbitals (see supplemental material D). Hopping parameters depend on distance between orbitals, and thus the same parameters are used in bilayer with and without strain, whatever the rotation angle. These parameters have been determined previously to simulate the dependence of van Hove singularities on the twist angle.\cite{Brihuega2012} In addition to providing structural information, the Fourier analysis outputs a commensurate approximate of the experimental structure which can be used for tight-binding calculations (See supplemental material E).  The top panel of Fig~\ref{f:tb}a displays the corresponding calculated LDOS. This calculation reproduces the three resonances seen in the measurements of Fig.~\ref{f:stm_sts}e. For comparison, Fig.~\ref{f:tb}b  shows the calculated LDOS of a similar commensurate approximate with the same rotation angle but exempt of heterostrain. In this case, the only two resonances are van Hove singularities arising from saddle points in the lowest energy bands.\cite{VanHove1953,LopesdosSantos2007,Li2009a,TramblydeLaissardiere2010,Bistritzer2011,LopesdosSantos2012,TramblydeLaissardiere2012,Brihuega2012,Wong2015} The saddle points marked by crosses in the central panel of Fig.~\ref{f:tb}b generate the positive energy singularity. The negative energy singularity is related to similar saddle points in the first negative energy band. The central panel of Fig.~\ref{f:tb}a shows that this band is completely reconstructed by heterostrain. The three-fold symmetry is lost and there are no longer Dirac points. Weakly dispersing regions appear around points $S''_1$ and $S''_2$ from which the zero energy resonance $E_0$ originates. The resonance at $E_1$ comes from the saddle points $S'_1$ and $S'_2$ and the resonance $E'_1$ comes from similar features in the first negative energy band. Cuts in the band structure presented in the lower panels of Fig.~\ref{f:tb}a and b demonstrate that the entire band structure is modified by the small heterostrain. Despite being undetectable by visual inspection of Fig.~\ref{f:stm_sts}d, heterostrain has nevertheless profound consequences. 

Interestingly, heterostrain does not suppress the electronic localisation induced by the moiré potential.\cite{TramblydeLaissardiere2010,Bistritzer2011,LopesdosSantos2012} Indeed, the central peaks in the LDOS are much more pronounced in AA regions than in AB regions in both experiments and calculations. Contrary to $E'_1$, $E_0$, $E_1$, the high energy resonances ($E_2$, $E_3$, $E_4$) in Fig.~\ref{f:stm_sts}e which are associated with a partial band gap opening in higher energy moiré bands\cite{Bistritzer2011,Wong2015,Kim2017a} are localised in AB regions. This finding which has never been reported is well reproduced by tight-binding. Further theoretical work is needed to understand its origin.

\begin{figure}
\includegraphics[width=0.4\textwidth]{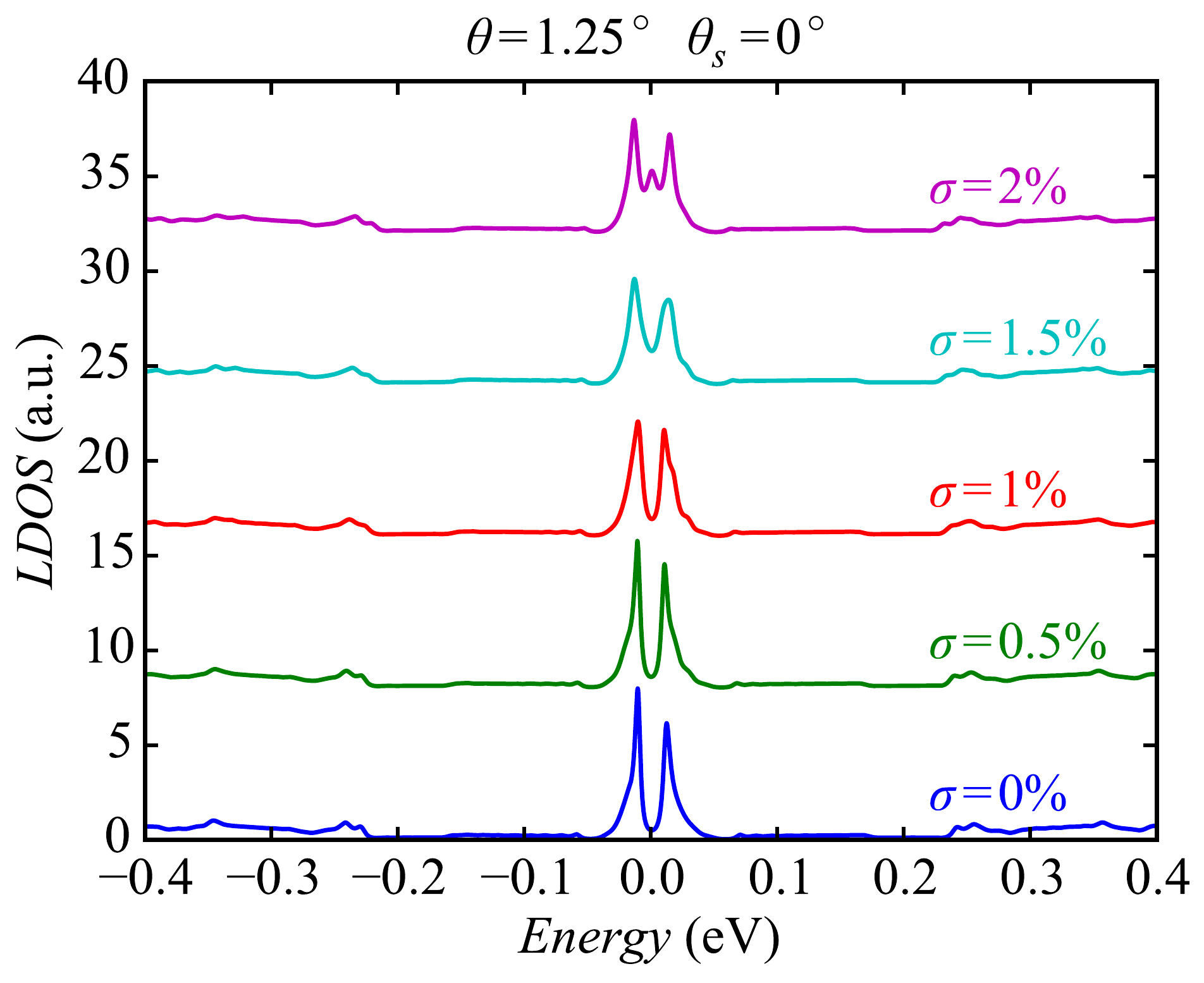}
\caption{\textbf{Calculated LDOS with homostrain.} Calculated LDOS in AA regions for the commensurate structure with $\theta=1.25^\circ$ for different uniaxial homostrain magnitudes along the zigzag direction of the bottom layer. All curves have been shifted vertically for clarity.}
\label{f:tb_dollfus}
\end{figure}

As a final verification, we have investigated the effect of homostrain since previous tight-binding calculations\cite{Nguyen2015} have shown that homostrain can also alter the band structure of TGLs with large rotation angles. For consistent comparison between homo- and heterostrain, we performed tight-binding calculations with 0.35~\% of uniaxial homostrain applied in the $x$ direction to the layers rotated by 1.25$^\circ$ (details of the calculations in supplemental material F). Figure~\ref{f:tb}c shows that in this case the LDOS is similar to the unstrained situation and the band structure is only weakly affected. This has to be expected since, as the layers are stretched together, the interlayer relative atomic positions evolve much slower than for heterostrain. This is why much larger values of homostrain (2\%) would be required to reproduce our experimental LDOS (see Fig.~\ref{f:tb_dollfus}). Such strain is much larger than observed by Raman in our sample (see supplemental material G) and usually reported for graphene on SiC.\cite{Beechem2014} As a consequence we exclude homostrain as the origin of our observations.

In conclusion, the properties of TGLs depend on the electronic coupling of the layers and hence on their relative arrangement. Heterostrain is therefore particularly efficient to tune their band structure. This should be the case of other homostructures (MoS$_2$, WS$_2$, WSe$_2$ etc.) and heterostructures((Gr/h-BN, MoSe$_2$/WSe$_2$ etc.) in which interlayer electronic also has a strong influence. In the future, a whole new generation of electronic devices could arise exploiting heterostrain in van der Waals stacks with intentionally individually-strained layers. It could prove instrumental in the exploration of the recently discovered strongly correlated electron physics in carbon materials.\cite{Hererro2018a, Hererro2018b}  

TLQ was supported by a CIBLE fellowship from Region Rhone-Alpes. We thank Valérie Reita and optics and microscopy technological group for valuable support on Raman spectroscopy. The authors wish to thank Johann Coraux for fruitful discussions.

\bibliography{Heterostrain}

\newpage
\section*{Supplemental material}
\renewcommand{\thefigure}{S\arabic{figure}}
\renewcommand{\theequation}{S\arabic{equation}}
\setcounter{figure}{0}
\begin{figure}[!htbp]
	\ind{5.5cm}{a} \includegraphics[width=0.45\textwidth]{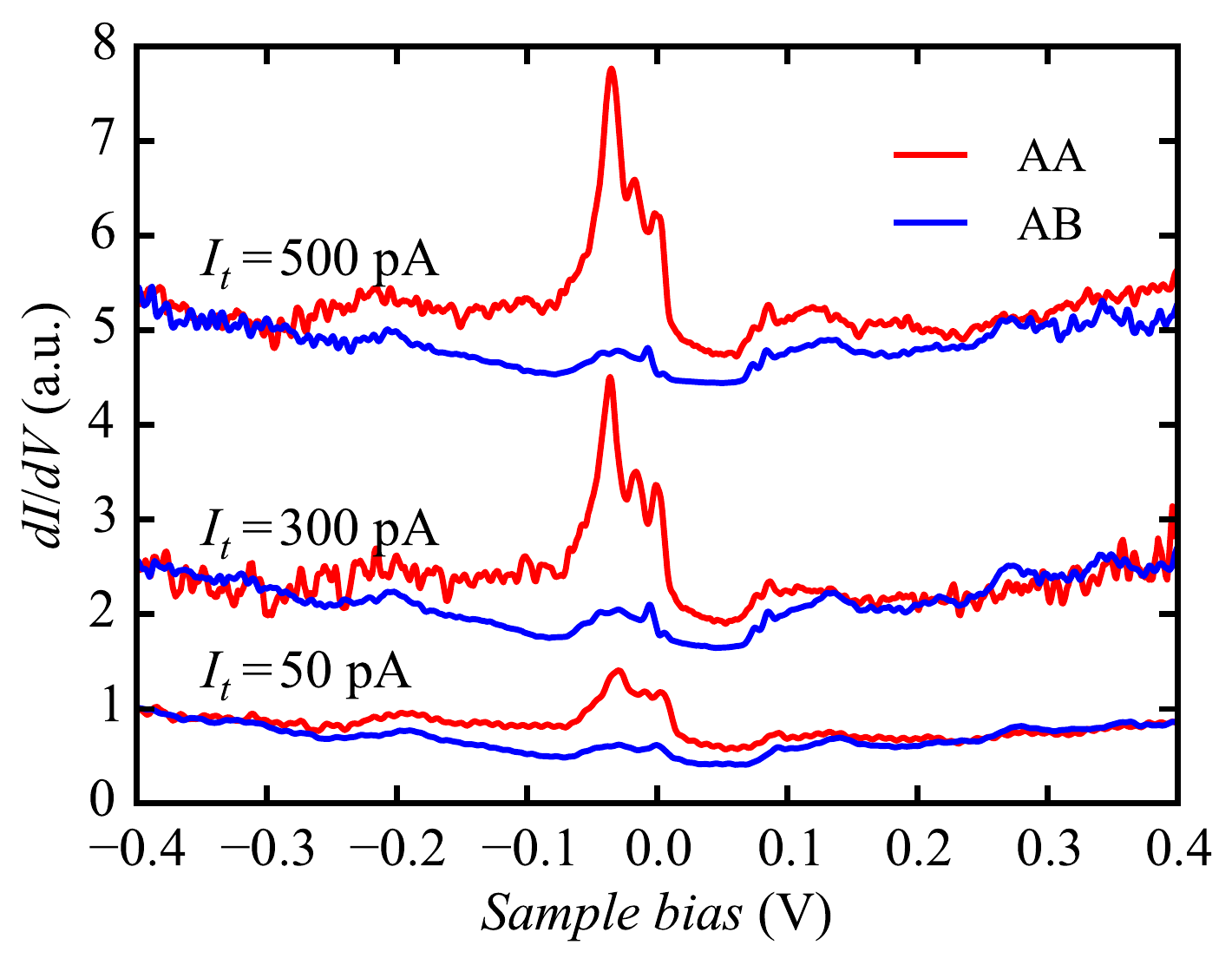}\\
	\includegraphics[width=0.45\textwidth]{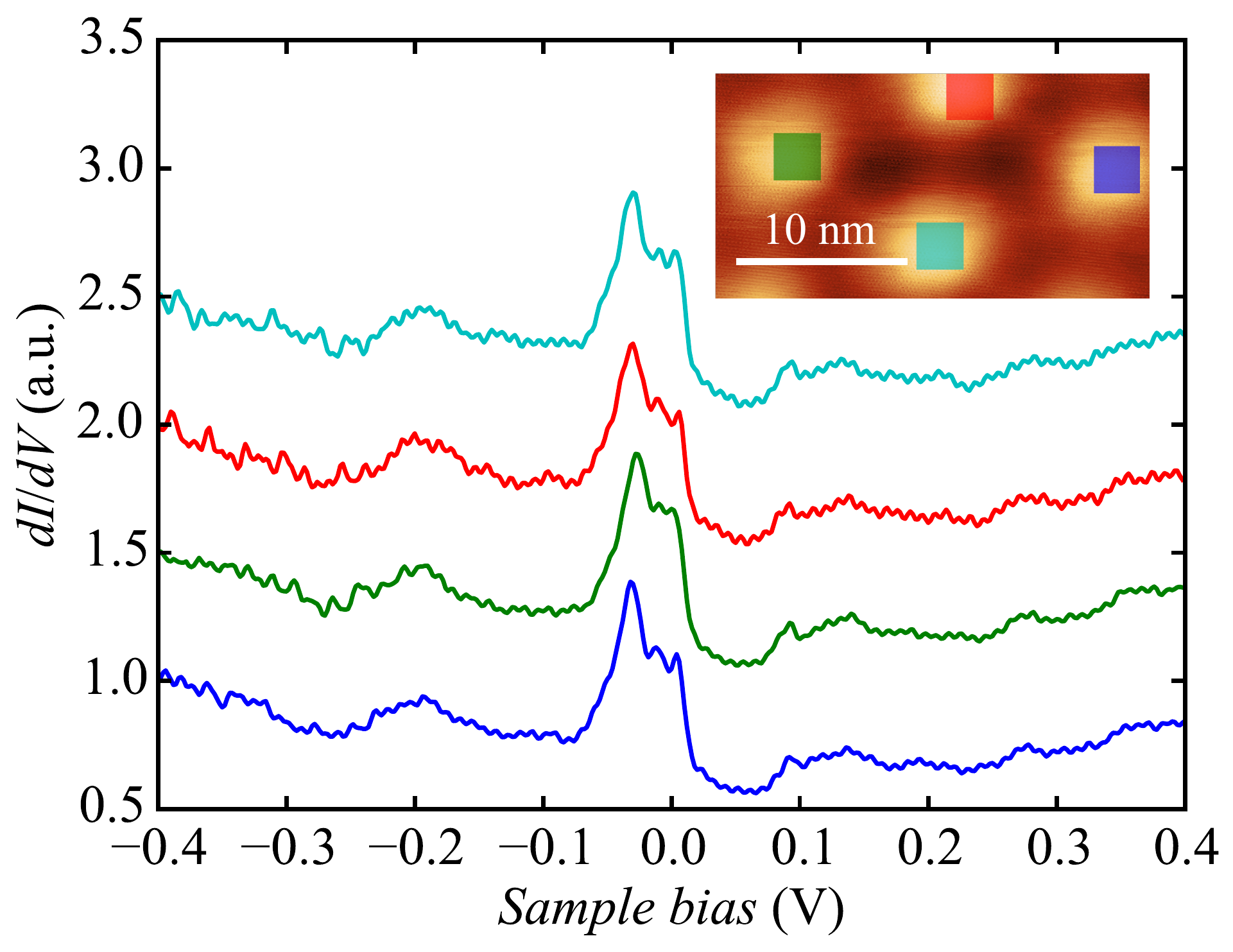}
	
	\caption{\textbf{Reproducibility and spatial variation of the DOS} a) Tunneling spectra measured in AA and AB regions with different tunneling conditions. b) Averaged STS spectra taken at different AA regions showing the reproducibility of the multiple vHs and the absence of significant difference between successive AA regions. Inset : (27.2 $\times$ 14) nm$^2$ STM topograph ($V=$\SI{-400}{\milli\volt},$I$=\SI{50}{\pico\ampere}) with colored rectangles taken for the area of averaging for the corresponding colored spectra. \label{f:reproductibility}}
\end{figure}

\subsection{Reproducibility of the data - Spatial variation of the LDOS}
Figure~\ref{f:reproductibility}a presents the local density of states in AA and AB regions as measured from $dI/dV$ spectra performed for different tunneling resistances. The spectra are very similar despite the tunneling resistance has been changed by an order of magnitude establishing the robustness of the experiment with respect to the tunneling conditions. We note that the data were normalized to the value at $V_b=-0.4$~V and the curves were shifted vertically for clarity. Figure~\ref{f:reproductibility}b Shows the spectroscopic measurements performed in different AA regions of the moiré (see inset for the location). This shows a very weak variability from neighboring AA region. The theoretical discussion of this absence of spatial variability is presented in section E (See Figure~\ref{f:moirons}).

\subsection{Fourier analysis}
\begin{figure}
	\ind{7.cm}{a}\includegraphics[width=0.95\columnwidth]{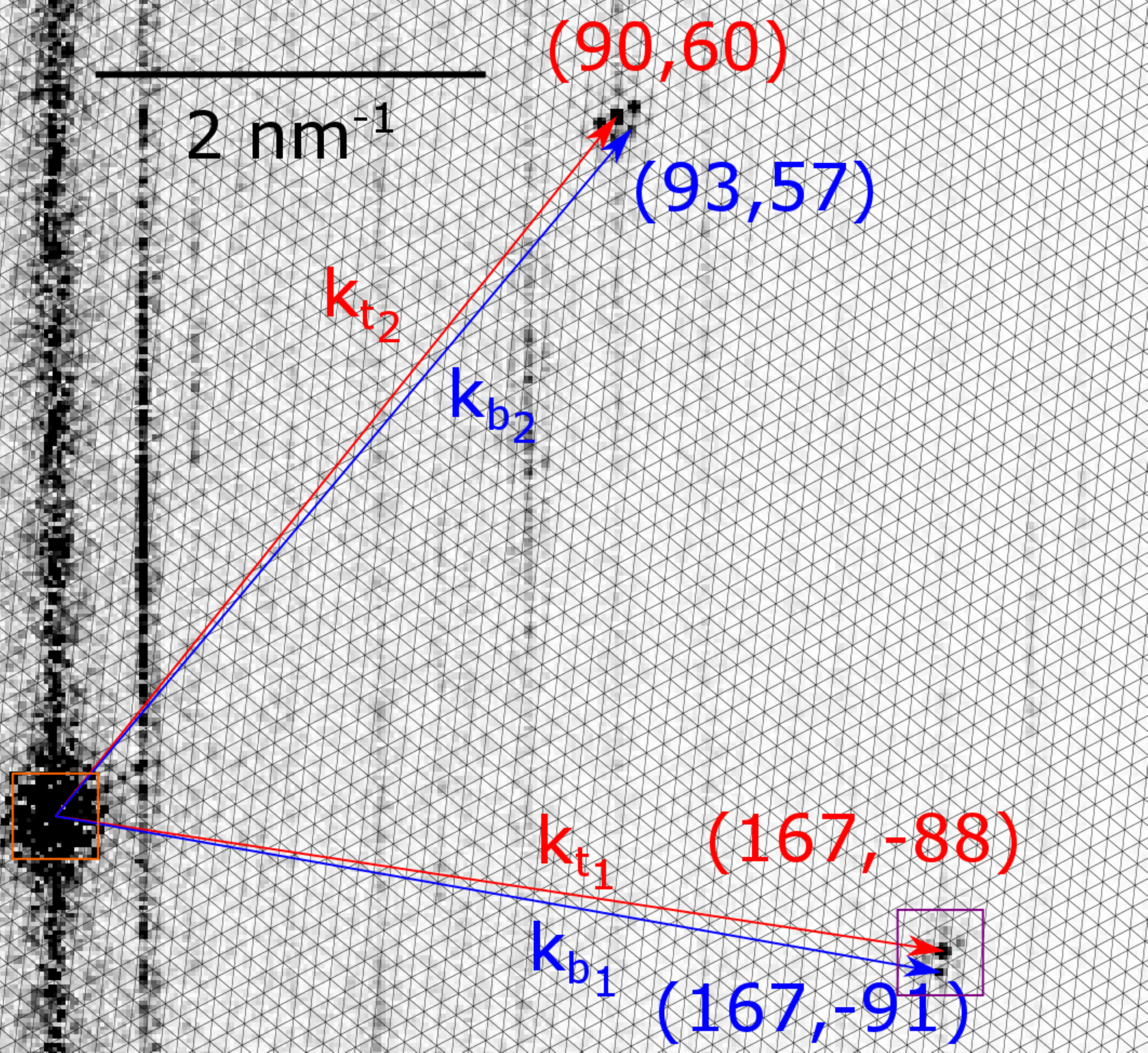} \\
	\ind{3.5cm}{b}\includegraphics[width=0.45\columnwidth]{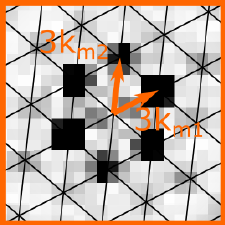}
	\ind{3.5cm}{c}\includegraphics[width=0.45\columnwidth]{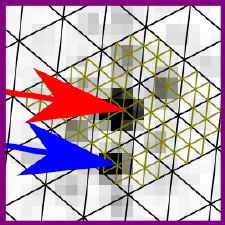}
	\caption{\textbf{Fourier analysis} a) Fourier transform (FT) of the STM topograph of the moiré pattern with atomic resolution. The black mesh decomposes the FT in the basis of the beating vectors $(3\vec{k_{m_1}},3\vec{k_{m_2}})$ (see text). b) Center of the FT (orange square) with enhanced contrast to see the decomposing vectors of the mesh $(3\vec{k_{m_1}},3\vec{k_{m_2}})$. c) Zoom on the graphene spots situated in the bottom right regions (purple square). The spots do not fall on the black beating mesh but instead on the green moiré mesh that is three times smaller.}
	\label{f:comm}
\end{figure}
\begin{figure*}
	\ind{4.5cm}{a}\includegraphics[height=5cm]{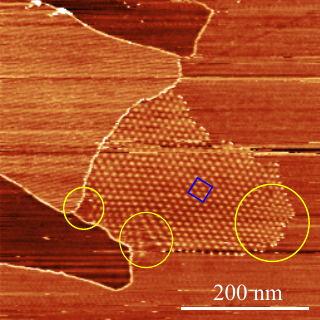} 
	\ind{4.5cm}{b}\includegraphics[height=5cm]{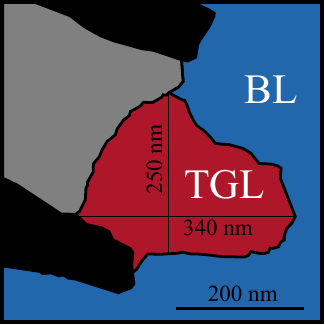}
	\caption{a) STM image of the domain of the probed moiré pattern ($V=$\SI{-500}{\milli\volt}, $I=$\SI{1}{\nano\ampere}). Deformation of the moiré is seen at the boundaries (yellow circles) where pinning of the top layer occurs. The STM and STS studies were done in the blue square where the moiré is regular, far from the boundaries  b) Color model showing the bottom layer (BL) in blue and the twisted graphene layers area (TGL) in red. The gray and black regions are other domains separated by grain boundaries.}
	\label{f:domain}
\end{figure*}
The Fourier analysis is a commensurability analysis which consists in finding coincidences of the atom positions between the graphene and its substrate, \cite{Artaud2016} in this case between the lattice vectors of the top graphene layer ($\vec{a_{t_1}}$,$\vec{a_{t_2}}$) and of the supporting bottom graphene layer ($\vec{a_{b_1}}$,$\vec{a_{b_2}}$). These coincidences can be written in a general form using the Park-Madden matrix:
\begin{equation}
\begin{pmatrix}
\vec{a_{t_1}} \\
\vec{a_{t_2}}
\end{pmatrix}
=
\begin{pmatrix}
A & B \\
C & D
\end{pmatrix}
\begin{pmatrix}
\vec{a_{b_1}} \\
\vec{a_{b_2}}
\end{pmatrix}
\label{e:pm_letters}
\end{equation}
where $(A,B,C,D)$ must be rational numbers ($\in\mathbb{Q}$) to achieve commensurability. \cite{Park1968} To find this decomposition from the experimental STM images, it is more accurate and practical to use reciprocal space vectors ($\vec{k_{t_1}}$,$\vec{k_{t_2}}$) and ($\vec{k_{b_1}}$,$\vec{k_{b_2}}$) that appear in the Fourier Transform (FT) of the images and to use the moiré vectors ($\vec{k_{m_1}}$,$\vec{k_{m_2}}$) as intermediates. The commensurability can be expressed from eight integers $(i,j,k,l,m,n,q,r)$ using
\begin{align} 
\vec{k_{t_1}}&=i\vec{k_{m_1}}+k\vec{k_{m_2}} \\
\vec{k_{t_2}}&=j\vec{k_{m_1}}+l\vec{k_{m_2}} \\
\vec{k_{b_1}}&=m\vec{k_{m_1}}+q\vec{k_{m_2}} \\
\vec{k_{b_2}}&=n\vec{k_{m_1}}+r\vec{k_{m_2}} \,.
\end{align}
From these eight integers, we get \cite{Artaud2016} 
\begin{equation}
\begin{pmatrix}
A & B \\
C & D
\end{pmatrix}
=
\dfrac{1}{il-jk}
\begin{pmatrix}
lm-jq & ln-jr \\
-km+iq & -kn+ir
\end{pmatrix}
\,.
\end{equation}
In the formalism of the extended Wood's notation ($P_1R\theta_1,P_2R\theta_2$), with heterostrain amplitudes $P_1=\dfrac{a_{t_1}}{a_{b_1}}$, $P_2=\dfrac{a_{t_2}}{a_{b_2}}$ and rotation angles $\theta_1=(\vec{a_{t_1}},\vec{a_{b_1}})$ and $\theta_2=(\vec{a_{t_2}},\vec{a_{b_2}})$,  the matrix linking ($\vec{a_{t_1}}$,$\vec{a_{t_2}}$) and ($\vec{a_{b_1}}$,$\vec{a_{b_2}}$) is given by
\footnotesize
\begin{equation}
\begin{pmatrix}
\vec{a_{t_1}} \\
\vec{a_{t_2}}
\end{pmatrix}
=
\begin{pmatrix}
P_1\left(\cos{\theta_1}+\dfrac{\sin{\theta_1}}{\sqrt{3}}\right) & \dfrac{2P_1}{\sqrt{3}}\sin{\theta_1} \\
-\dfrac{2P_2}{\sqrt{3}}\sin{\theta_2} & P_2\left(\cos{\theta_2}-\dfrac{\sin{\theta_2}}{\sqrt{3}}\right)
\end{pmatrix}
\begin{pmatrix}
\vec{a_{b_1}} \\
\vec{a_{b_2}}
\end{pmatrix}
\end{equation}
\normalsize
when considering a hexagonal support lattice. The parameters $P_1$, $P_2$, $\theta_1$ and $\theta_2$ can be found from the commensurability by identifying this matrix with the Park-Madden matrix defined in \eqref{e:pm_letters}:
\begin{align}
P_1=\sqrt{A^2+B^2-AB} \\
P_2=\sqrt{C^2+D^2-CD} \\
\theta_1=\arctan{\dfrac{B\sqrt{3}}{2A-B}} \\
\theta_2=\arctan{\dfrac{C\sqrt{3}}{2C-D}}
\end{align}
Uniaxial and biaxial heterostrains levels $\varepsilon_{uni}^{het}$ and $\varepsilon_{bi}^{het}$ are then given by
\begin{align}
\varepsilon^{het}_{bi}=&\sqrt{X-\sqrt{W}}-1 \\
\varepsilon^{het}_{uni}=&\sqrt{X+\sqrt{W}}-\sqrt{X-\sqrt{W}}
\end{align}
where $X=\dfrac{2(P_1^2+P_2^2)+A}{3}$, $Y=\sqrt{\dfrac{4P_1^2P_2^2-A^2}{3}}$ and $W=X^2-Y^2$. The detailed derivation can be found in the supplementary of Ref.~\onlinecite{Artaud2016}.

To find the commensurability indices from which every structural parameter stems, we decomposed the Fourier transform (FT) shown in Fig.~\ref{f:comm}a in the basis of the moiré beating vectors $\vec{k_{t_1}} - \vec{k_{b_1}}$ and $\vec{k_{t_2}} - \vec{k_{b_2}}$. As underlined in Refs.~\onlinecite{Artaud2016} and ~\onlinecite{Zeller2017}, the beating periodicity, that is the distance between two bright spots, is not necessarily the true periodicity of the moiré system which can span several beatings. In this case, the graphene reciprocal vectors cannot be decomposed in integer coordinates in the mesh of the beating period. Indeed, closer inspection in Fig~\ref{f:comm}c shows that the graphene spots are at one third of the mesh which means that the moiré reciprocal vectors are three times smaller than the beating reciprocal vectors. Therefore, in direct space, the moiré system spans three inequivalent beatings in each direction. By defining $3\vec{k_{m_1}}=\vec{k_{t_1}} - \vec{k_{b_1}}$ and $3\vec{k_{m_2}}=\vec{k_{t_2}} - \vec{k_{b_2}}$, the commensurate structure is defined by
\begin{align}
\vec{k_{t_1}}&=167\vec{k_{m_1}}-88\vec{k_{m_1}} \\
\vec{k_{t_2}}&=90\vec{k_{m_1}}+60\vec{k_{m_2}} \\
\vec{k_{b_1}}&=167\vec{k_{m_1}}-91\vec{k_{m_2}} \\
\vec{k_{b_2}}&=93\vec{k_{m_1}}+57\vec{k_{m_2}}\,.
\end{align}
At this stage, the commensurability relations can be written using the previously introduced Park-Madden matrix:
\begin{equation}
\begin{pmatrix}
\vec{a_{t_1}} \\
\vec{a_{t_2}}
\end{pmatrix}
=
\dfrac{1}{17940}
\begin{pmatrix}
18210 & 450 \\
-510 & 17703
\end{pmatrix}
\begin{pmatrix}
\vec{a_{b_1}} \\
\vec{a_{b_2}}
\end{pmatrix}
\label{e:pm}\,.
\end{equation}
Once these relations are known, we are then able to extract the parameters of the extended Wood's relation and the heterostrains $\varepsilon_{bi}^{het}$ and $\varepsilon_{uni}^{het}$ from the equations given above.
For the present structure, we find ($P_1R{\theta_1}\times P_2R{\theta_2}$) = ($1.00274R1.241\times 1.00104R1.381$), $\varepsilon_{uni}^{het}=0.35\% (\pm 0.03\%)$ and $\varepsilon_{bi}^{het}=-0.06\% (\pm 0.005\%)$ (angles are given in degree). In the following this will be referred to as \textbf{structure 1}.

Additionally, we note that since experimental FT have a finite resolution, it is always possible to fit the atomic lattice spots by dividing the beating mesh by an integer value. The commensurability hypothesis is in this framework always fulfilled. The analysis therefore yields a commensurate approximate which may not be the actual structure but is very close to it. The commensurate approximate allows to find the structural parameters (angle and strain) with sufficient accuracy. 

\subsection{Origin of heterostrain}

As suggested in the main text, heterostrain between the layers originates from the finite extension of the top layer as supported by the lattice variations magnified by the moiré near the grain boundary in Figure~\ref{f:domain}. The STS measurements described in the main text were performed in the blue square shown in Fig.~\ref{f:domain}a. This region sits at the center of the graphene grain where the moiré pattern appears (TGL region in Fig.~\ref{f:domain}b). The variations in the moiré pattern near the edges of this grain (yellow circles in Fig.~\ref{f:domain}a) provides evidence that strain is applied at the grain boundary where the top layer is most probably pinned. Indeed, since the moiré results from an interference effect, any deformation of the atomic lattice is magnified by the moiré pattern. The bottom layer being much more extended (spanning at least the blue and red region in Fig. ~\ref{f:domain}b), it is not affected by this local strain or to a much lesser extent. Such a dissymmetry spontaneously induces a heterostrain between the two layers. Away from the boundary, the moiré pattern becomes regular indicating that this heterostrain is uniform and has reached the measured values.

\subsection{Tight-binding calculations}
In this section, we present the tight-binding scheme that was also used in previous works \cite{TramblydeLaissardiere2010,TramblydeLaissardiere2012,Brihuega2012,Cherkez2015}. The same tight-binding parameters are used to compute local density of states (LDOS) for all presented calculations. It has been shown that these parameters simulated accurately experimental measurements of the van Hove singularities for twisted graphene layers with a rotation angle between $2^{\rm o}$ and $20^{\rm o}$ \cite{Brihuega2012,Cherkez2015}.

Only $p_z$ orbitals of C atoms are taken into account since we are interested in what happens around the Fermi level. In rotated bilayers interlayer interactions between two $p_z$ orbitals are thus not restricted to $pp\sigma$ terms but $pp\pi$ terms have also to be introduced. The Hamiltonian has the form:
\begin{equation}
H = \sum_{i} \epsilon_i \, |i\rangle \langle i|  
~+~  \sum_{<i,j>} t_{ij} \, |i\rangle \langle j|
\label{eq:hamilt}
\end{equation}
where  $|i\rangle$ is the $p_z$ orbital located at $\vec r_i$, and $\langle i,j\rangle$ is the sum on index $i$ and $j$ with $i\ne j$. The coupling matrix element, $t_{ij}$, between two $p_z$ orbitals located at $\vec r_i$ and $\vec r_j$ is \cite{Slater1954}
\begin{equation}
t_{ij} 
~=~   n^2 V_{pp\sigma}(r_{ij}) ~+~ (1 - n^2) V_{pp\pi}(r_{ij})
\end{equation}
where $n$ is the direction cosine of $\vec r_{ij} = \vec r_j - \vec r_i$ along O$z$ axis perpendicular to graphene planes and $r_{ij}$ the distance $r_{ij}$ between the orbitals:
\begin{equation}
n ~=~ \frac{z_{ij}}{r_{ij}} ~~~{\rm and}~~r_{ij} = \| \vec r_{ij} \| .
\end{equation}
$z_{ij}$ is the coordinate of $\vec r_{ij}$ along O$z$. It is either equal to zero or to a constant because the two graphene layers have been kept flat in our model.
We use the same following dependence on distance of the Slater-Koster parameters:
\begin{align}
V_{pp\pi}(r_{ij})    &=-\gamma_0 \,{\rm e}^{q_{\pi}    \left(1-\frac{r_{ij}}{a_0} \right)} , \label{eq:tb0}\\
V_{pp\sigma}(r_{ij}) &= \gamma_1 \,{\rm e}^{q_{\sigma} \left(1-\frac{r_{ij}}{a_1} \right)}
\label{eq:tb1}
\end{align}
where $a_0=1.418$~{\rm \AA} is the nearest neighbor distance within a layer without strain, and $a_1=3.349$~{\rm \AA} is the interlayer distance.

First neighbors interaction in a plane $\gamma_0$ is set to have the Fermi velocity in graphene equal to $1.09$$\times$$10^6$~m.s$^{-1}$, $\gamma_0 = 3.7$~eV \cite{Brihuega2012}. Second neighbors interaction $\gamma_0'$ in a plane is set  to $0.1\gamma_0$ \cite{CastroNeto2009} which fixes the value of the ratio $q_{\pi}/{a_0}$ in equation (\ref{eq:tb0}).The inter-layer coupling between two $p_z$ orbitals in $\pi$ configuration is $\gamma_1$. $\gamma_1$ is fixed to obtain a good fit with DFT calculation around Dirac energy in AA stacking and AB stacking which gives $\gamma_1=0.48$~eV.
To get $q_{\pi}$, we have chosen the same coefficient of the exponential decay for $V_{pp\pi}$ and $V_{pp\sigma}$,
\begin{equation}
\frac{q_{\sigma}}{a_1} ~=~ \frac{q_\pi}{a_0} 
~=~ \frac{{\ln} \left(\gamma_0/\gamma_0' \right)}{a - a_0}
~=~ 2.218  \,{\rm \AA}^{-1},
\end{equation}
with $a = \sqrt{3} a_0 =  2.456$\,{\rm \AA} the distance between second neighbors in a plane without strain. All $p_z$ orbitals have the same on-site energy $\epsilon_i$ (Eq.~\ref{eq:hamilt}). $\epsilon_i$ is set to $-1.0736$\,eV so that the energy $E_{\rm D}$ of the Dirac point is equal to zero in mono-layer graphene without strain. $\epsilon_i$ is not zero because the intra-layer coupling between atoms beyond first neighbors breaks the electron/hole symmetry and then shifts $E_{\rm D}$.
\begin{figure}
	\includegraphics[width=\columnwidth]{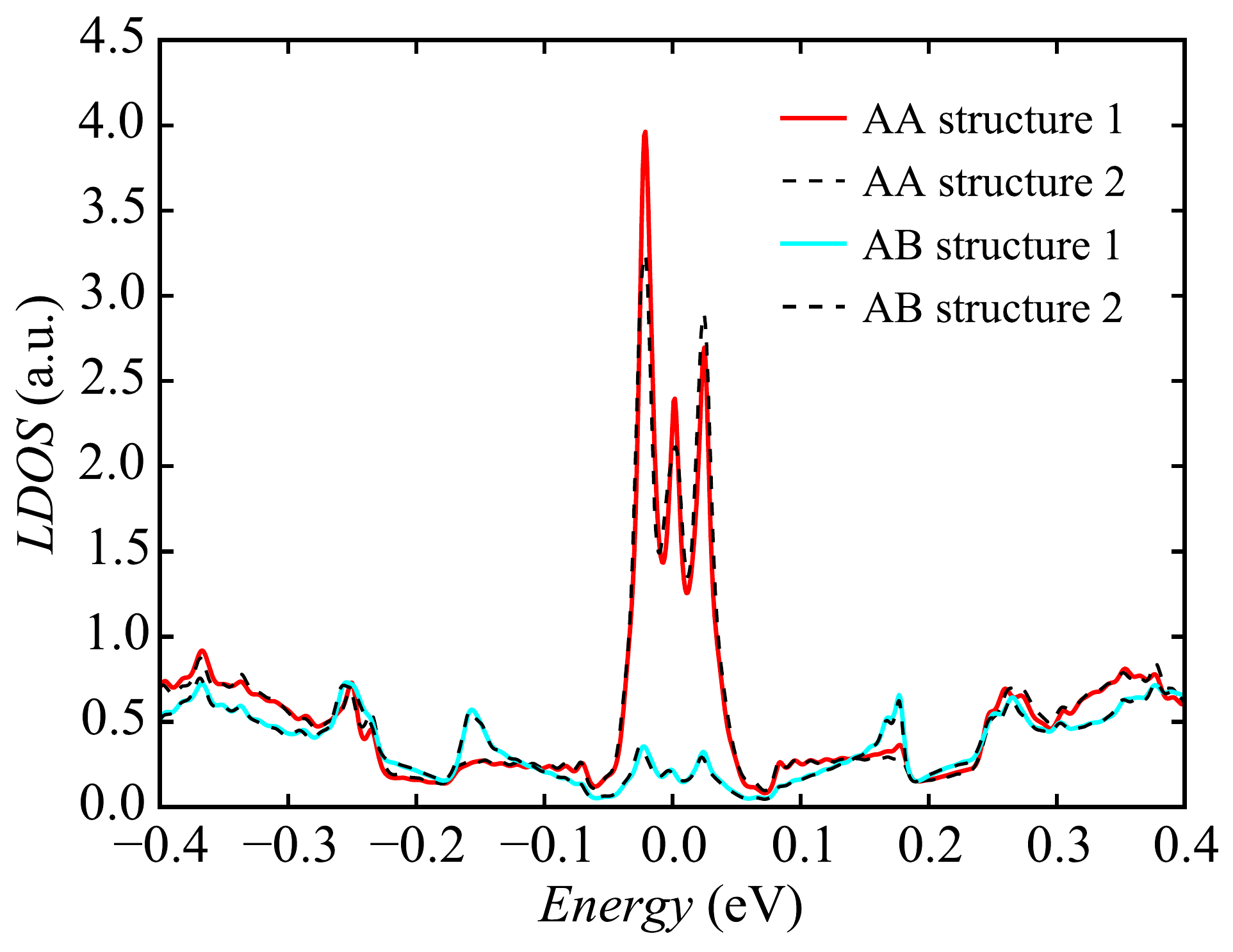}
	\caption{\textbf{Comparison of structure 1 and 2}. LDOS calculated for \textbf{structure 1} and \textbf{structure 2} in AA and AB regions. The two approximations are very close to each other.}
	\label{f:Dos_compare}
\end{figure}
LDOS on a $p_z$ orbitals are computed for periodic bilayers containing a very large number of atoms in a unit cell. Numerical diagonalization of the Hamiltonian is thus impossible so we have used the recursion method in real space based on Lanczos algorithm \cite{Haydock1972}. To reach a precision in energy of about $0.2$~meV, we have computed 3000 steps in the continued fraction built in a super-cell of more than 12 millions of orbitals, with periodic boundary conditions.

\begin{figure}[!htbp]
	\ind{5.5cm}{a} \includegraphics[width=0.45\textwidth]{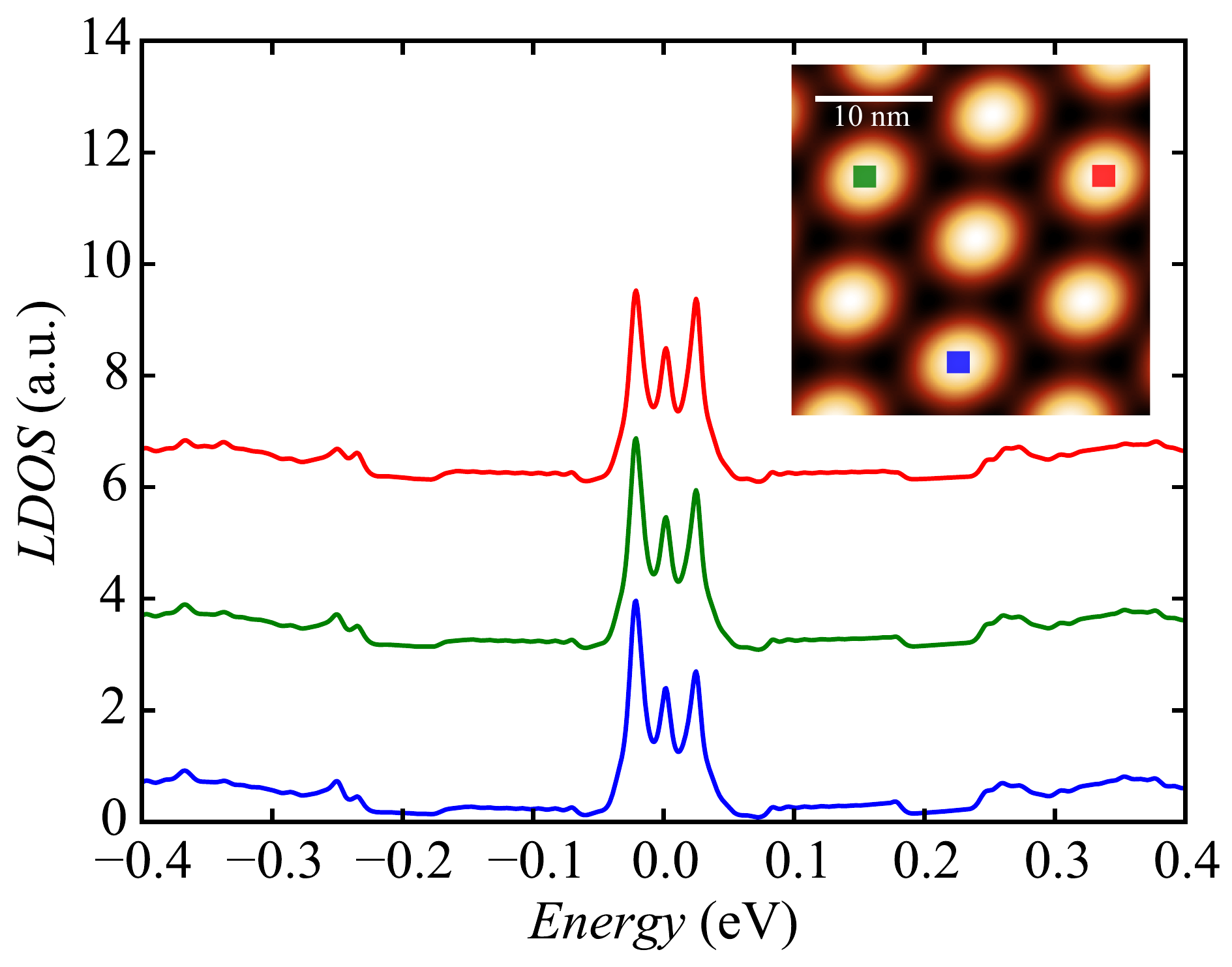}\\
	
	\caption{\textbf{Spatial variation of the DOS in AA regions} Tight-binding calculation of the LDOS at different AA regions of the moiré cell. Only very slight modulations of the amplitude of the vHs are seen. Inset : moiré pattern generated from the commensurability with colored rectangles indicating the location at which the LDOS were calculated.  \label{f:moirons}}
\end{figure}

\begin{figure*}
	\includegraphics[angle=270]{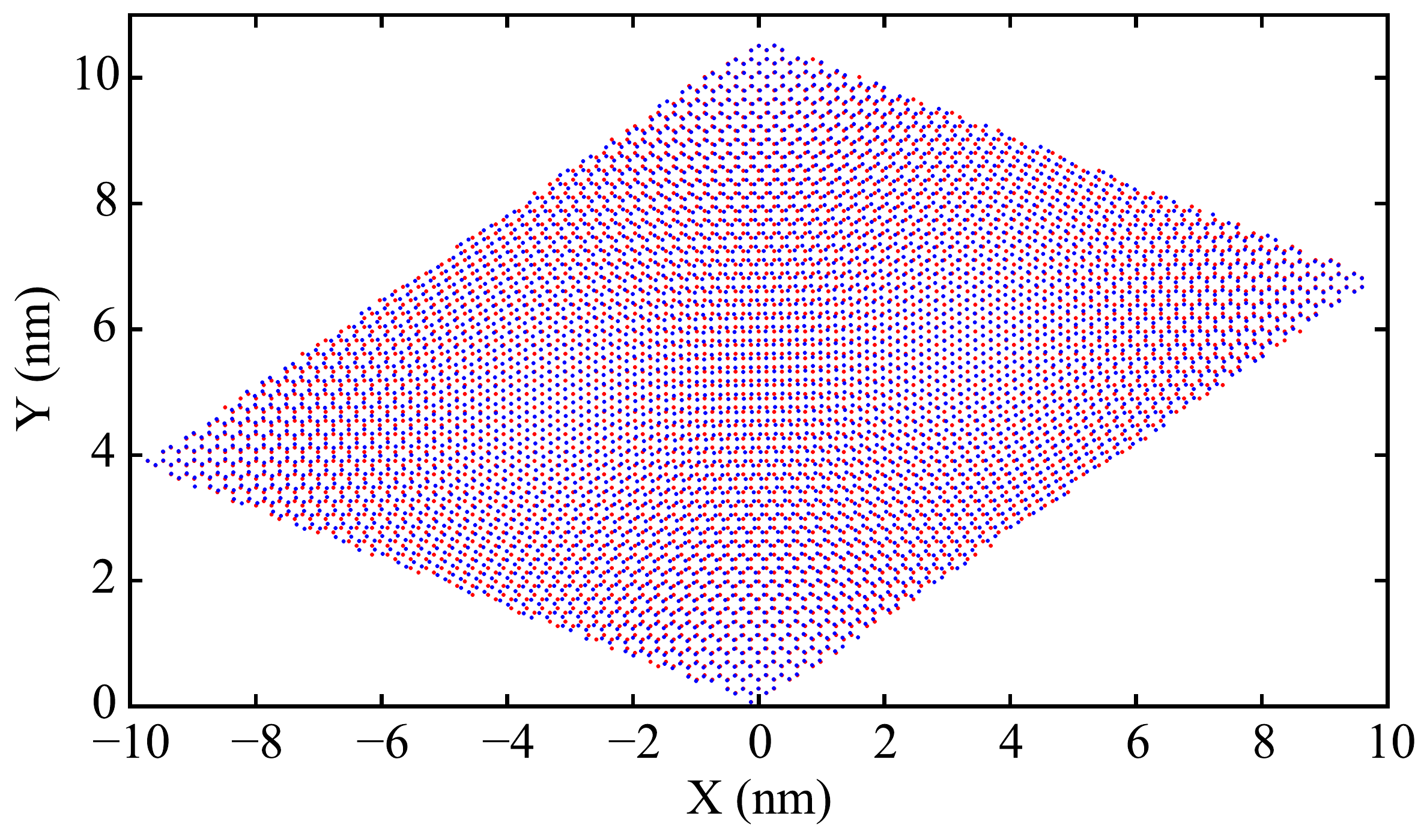}
	\caption{\textbf{Moiré cell.} Generated atomic positions for the tight-binding calculations with \textbf{structure 2} (red points: top layer, blue points: bottom layer). .}
	\label{f:struct}
\end{figure*}
\subsection{Generation of the structure for tight-binding calculations}
The positions of the atoms for the tight-binding calculations were generated from the commensurability relations so that the heterostrain is inherently taken into account. Once the atomic positions of the bottom layer are defined, those of the top layer are obtained directly from Eq.~\eqref{e:pm}. For the calculations in the main text, we assumed that the bottom layer is unstrained. The periodic cell generated from \textbf{structure 1} contains 71844 atoms. 
While the calculation of the LDOS is possible for such large system, calculation of its band structure would take unrealistic time .
As a consequence we decided to work with a smaller size commensurate approximate (\textbf{structure 2}) which in extended Wood's notation reads ($1.00324R1.24\times  1.0013R1.392$) and contains only 7988 atoms. In this approximate the heterostrain is $\varepsilon_{uni}^{het}=0.38\%$ and $\varepsilon_{bi}^{het}=-0.04\%$. Its unit cell is presented in Fig.~\ref{f:struct}.

Figure~\ref{f:Dos_compare} demonstrates that the two structures have a LDOS very close to each other. It is worth noting that the two structures differ in that \textbf{structure 1} is a high order commensurate structure while \textbf{structure 2} is a first order one: in direct space the moiré of \textbf{structure 1} spans three inequivalent beatings in each direction while that of \textbf{structure 2} spans only one.

Figure~\ref{f:moirons}, shows that while the atomic arrangements in adjacent moiré beating are not strictly the same in \textbf{structure 1} the calculated LDOS is very weakly affected by those differences. As a consequence, neglecting these differences by using \textbf{structure 2} in which all beatings are equivalent is appropriate. This is supported by experiments since no significant variations of the LDOS is seen (Fig.~\ref{f:reproductibility}b). The absence of any meaningful difference in the LDOS is also in agreement with the continuum model of Ref.~\onlinecite{LopesdosSantos2012} where the authors concluded that a multiple beatings cell is a quasi-periodic repetition of a single beating cell.

\begin{figure}
	\includegraphics[width=0.415\textwidth]{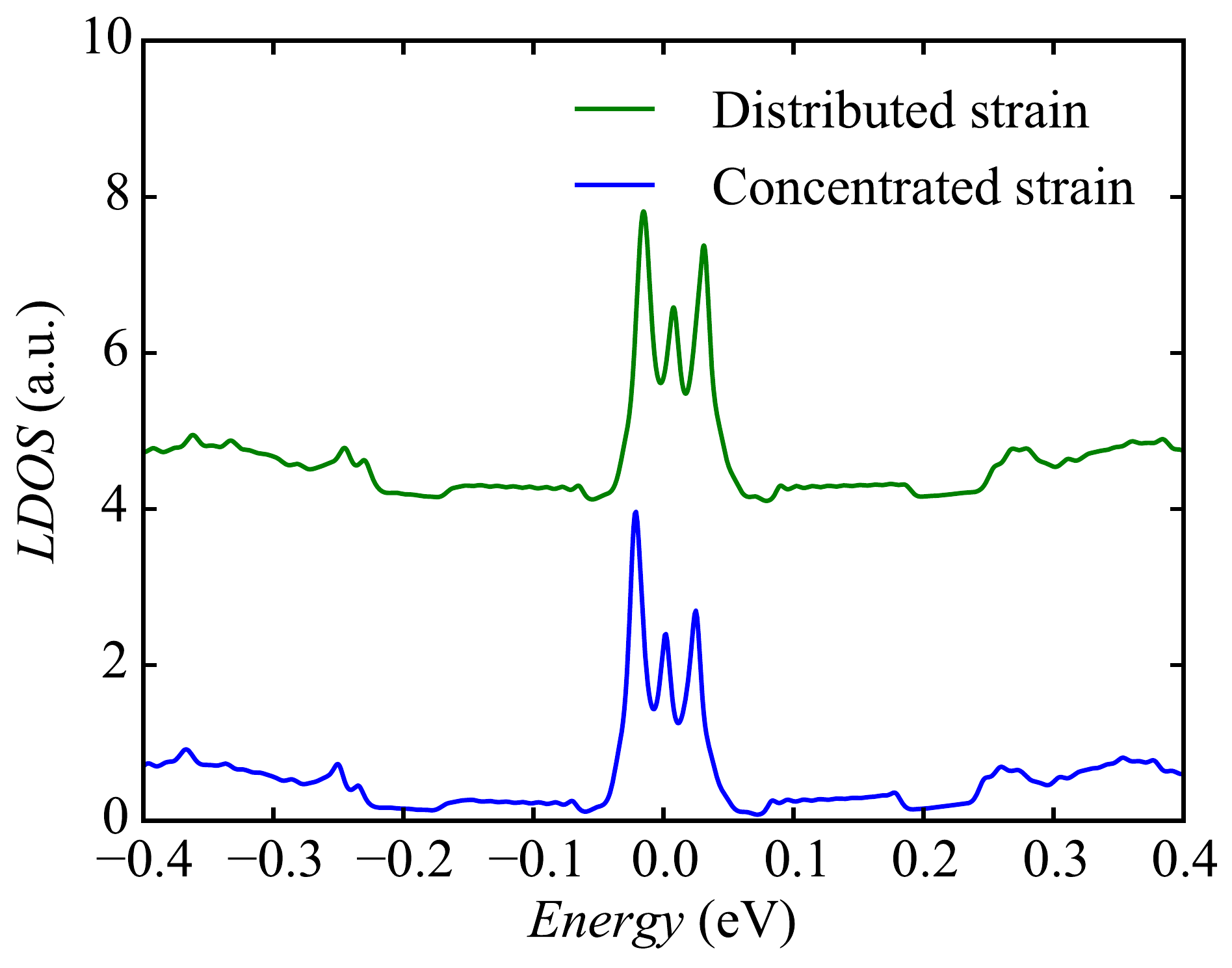}
	\caption{Local density of states in AA regions of the top layer calculated by tight-binding for the case where all the strains are concentrated in the top layer and for the case where the uniaxial strain was distributed in the two layers. Heterostrain between the layers is the same as the commensurability relations stay valid which explains the absence of change in the LDOS. The green curve has been shifted for clarity.}
	\label{f:tb_distrib}
\end{figure}

We have then investigated the situation where strain is distributed in the two layers. For this purpose, we applied half of the uniaxial strain to the atomic positions of the bottom layer and determined those of the top layer through the commensurability relations in order to keep heterostrain unchanged. the calculations were done using \textbf{structure 1}. Figure~\ref{f:tb_distrib} shows the tight-binding calculations of the LDOS in AA regions for both situations, where heterostrain is either confined in one layer or distributed between the two layers. The two results are very similar, with three resonances at the same energies, confirming the robustness of our model to this assumption.

\subsection{Effect of homostrain}
We detail here the procedure used to get the tight-binding results on TGL with homostrain presented in Fig.~3 of the main text.
Similarly to Ref.~\onlinecite{Nguyen2015} the effect of an uniaxial homostrain was evaluated by applying the following strain matrix to the atomic structure of the pristine TGL:
\begin{equation}
I_2+
\sigma
\begin{pmatrix}
\cos^2\theta_s-\gamma\sin^2\theta_s & (1+\gamma)\cos\theta_s\sin\theta_s \\ (1+\gamma)\cos\theta_s\sin\theta_s & \sin^2\theta_s-\gamma\cos^2\theta_s
\end{pmatrix}
\end{equation}
where $I_2=\begin{pmatrix}
1&0 \\
0&1
\end{pmatrix}$ is the two-dimensional identity matrix, $\sigma$ the strain amplitude, $\theta_s$ the angle with respect to the first crystallographic direction of the bottom layer ($\vec{a_{b_1}}$) and $\gamma$ the Poisson ratio set to $0.165$. \cite{Blakslee1970} This matrix is the result of an uniaxial strain matrix of magnitude $\sigma$ along $\theta_s$ transposed in the basis of the layers. Note that the convention for defining $\theta_s$ is different from the one used in Ref.~\onlinecite{Nguyen2015}. The resulting structure was used to perform tight-binding calculation of the LDOS in AA regions as shown in Fig.~3 of the main text.
\begin{figure}
	\includegraphics[width=0.415\textwidth]{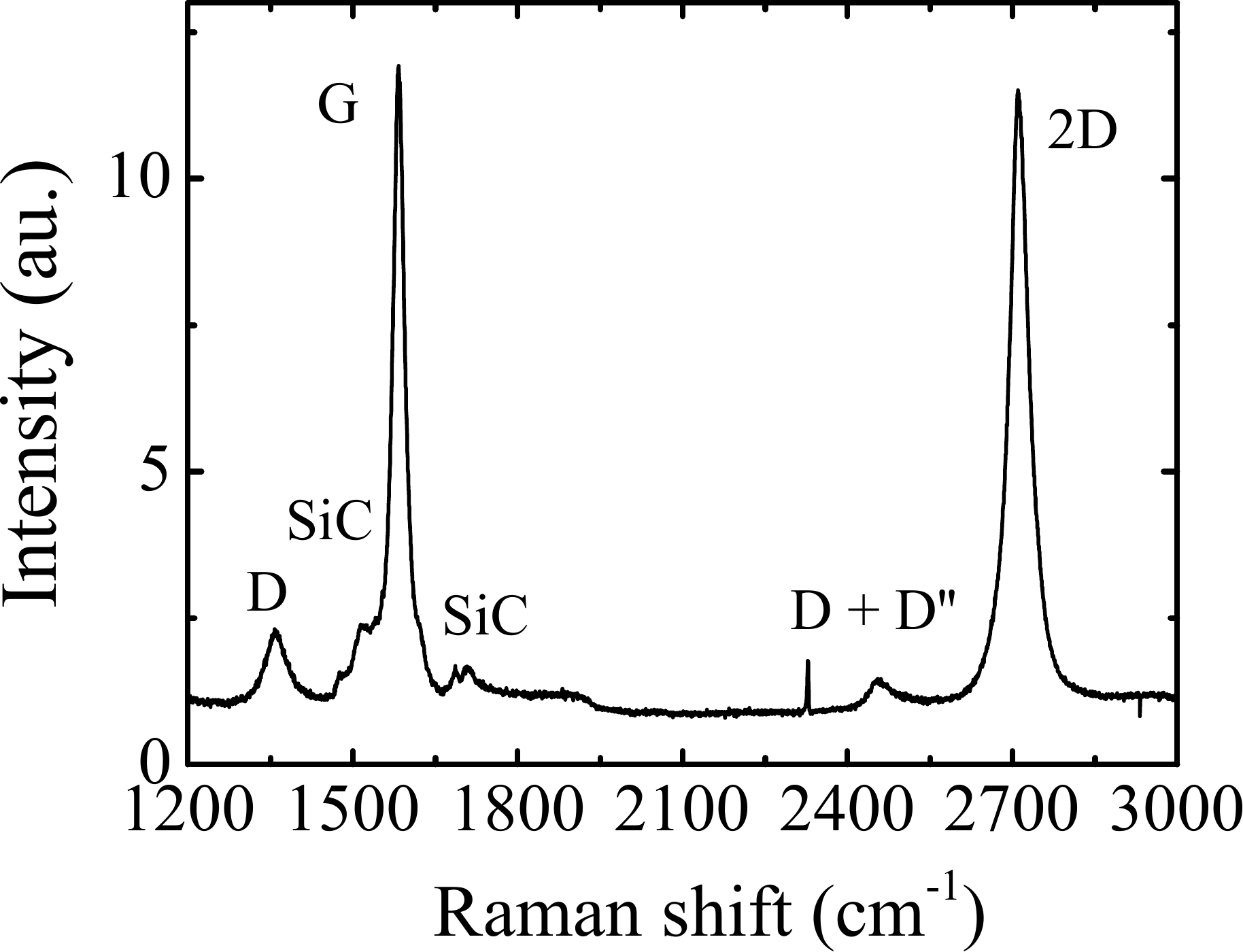}
	\caption{Raman spectrum of graphene grown on SiC(000$\bar{1}$) showing the typical G,D and 2D peaks. The position of the G peak indicates a strain value smaller than 0.2\%.}
	\label{f:Raman}
\end{figure}

\subsection{Raman spectroscopy}
Our typical Raman measurements presented in Fig.~\ref{f:Raman} also support that homostrain is small. The position of the G peak is observed at 1587.9 cm$^{-1}$ and the 2D peak at 2711.3 cm$^{-1}$. In rotated graphene layers, the position of the 2D band is difficult to interpret as the 2D Raman process depends strongly on the electronic dispersion. \cite{Ferrari2013} The position of the G peak is more straightforward to analyze since it is not affected by the rotation angle \cite{Havener2012} but only by doping and strain. The STS spectra in the main text show a n-doping of 25 meV which is too small to have any visible effect on the Raman spectrum. \cite{Das2008a} The G peak in our sample is therefore only affected by uniaxial strain which is supposed to split the G peak for strains higher than 0.2\%. Since we do not observe such a splitting, 
the strain in our sample has to be smaller than 0.2\%, in agreement with previous measurements. \cite{Beechem2014} \\
For the Raman measurements, a circularly polarized Ar-laser ($\lambda = 514.5$~nm) with laser power $\sim1$~mW was used. The laser beam was focused onto the sample using a Olympus 100 microscope objective (0.9 numerical aperture), which was also used to collect the scattered light. This scattered light was dispersed by a Jobin-Yvon T64000 single stage micro-Raman spectrometer (1800~grooves/mm), and collected by a liquid-nitrogen-cooled charge-coupled-device (CCD) detector. The spatial resolution was less than 1~$\mu$m, and the spectral resolution was about 1~cm$^{-1}$.

\bibliography{Heterostrain}

\end{document}